\documentstyle[12pt,a4wide]{article}
\newcommand{\be}{\begin{equation}}
\newcommand{\ee}{\end{equation}}
\newcommand{\bea}{\begin{eqnarray}}
\newcommand{\eea}{\end{eqnarray}}
\newcommand{\bean}{\begin{eqnarray*}}

\newcommand{\eean}{\end{eqnarray*}}
\font\upright=cmu10 scaled\magstep1
\font\sans=cmss10
\newcommand{\ssf}{\sans}
\newcommand{\stroke}{\vrule height8pt width0.4pt depth-0.1pt}
\newcommand{\Z}{\hbox{\upright\rlap{\ssf Z}\kern 2.7pt {\ssf Z}}}

\newcommand{\C}{{\rlap{\rlap{C}\kern 3.8pt\stroke}\phantom{C}}}
\newcommand{\R}{\hbox{\upright\rlap{I}\kern 1.7pt R}}
\newcommand{\CP}{\C{\upright\rlap{I}\kern 1.5pt P}}
\newcommand{\PP}{\hbox{\upright\rlap{I}\kern 1.5pt P}}

\newcommand{\identity}{{\upright\rlap{1}\kern 2.0pt 1}}

\newcommand{\HH}{\mbox{\hbox{\upright\rlap{I}\kern 1.7pt H}}}

\newcommand{\ci}{{\cal I}}
\newcommand{\I}{{\cal I}}
\def\dalemb#1#2{{\vbox{\hrule height .#2pt
\hbox{\vrule width.#2pt height#1pt \kern#1pt\vrule width.#2pt}
\hrule height.#2pt}}}
\def\square{\mathord{\dalemb{5.9}{6}\hbox{\hskip1pt}}}

\font\mybb=msbm10 at 11pt

\def\bb#1{\hbox{\mybb#1}}

\def\bC {\bb{C}}

\renewcommand{\CP}{\bC {\rm P}}

\newcommand{\news}{\setcounter{equation}{0}}
\input{epsf}
\begin{document}
\title{\vskip -70pt
\begin{flushright}
\end{flushright}\vskip 50pt
{\bf \large \bf SKYRMIONS, FULLERENES AND RATIONAL MAPS }\\[30pt]
\author{Richard A. Battye$^{\ \dagger}$ and Paul M. Sutcliffe$^{\ \ddagger}$
\\[10pt]
\\{\normalsize $\dagger$ {\sl Department of Applied Mathematics and Theoretical
Physics,}}
\\{\normalsize {\sl Centre for Mathematical Sciences, University of Cambridge,}}
\\{\normalsize {\sl Wilberforce Road, Cambridge CB3 0WA, U.K.}}
\\{\normalsize {\sl Email : R.A.Battye@damtp.cam.ac.uk}}\\
\\{\normalsize $\ddagger$  {\sl Institute of Mathematics, University of Kent at Canterbury,}}\\
{\normalsize {\sl Canterbury, CT2 7NZ, U.K.}}\\
{\normalsize{\sl Email : P.M.Sutcliffe@ukc.ac.uk}}\\}}
\date{March 2001}
\maketitle

\begin{abstract}
We apply two very different approaches to calculate Skyrmions with
baryon number $B\le 22.$  The first employs the rational map ansatz,
where approximate charge $B$ Skyrmions are constructed from a degree
$B$ rational map between Riemann spheres. We use a simulated annealing
algorithm to search for the minimal energy rational map of a given
degree $B.$ The second involves the numerical solution of the full
non-linear time dependent equations of motion, with initial conditions
consisting of a number of well separated Skyrmion clusters. In
general, we find a good agreement between the two  approaches.  For
$B\ge 7$ almost all the solutions are of fullerene type, that is, the
baryon density isosurface consists of twelve pentagons and $2B-14$
hexagons arranged in a trivalent polyhedron.  There are exceptional
cases where this structure is modified, which we discuss in detail. We
find that for a given value of $B$ there are often many Skyrmions,
with different symmetries, whose  energies are very close to the
minimal value, some of which we discuss.  We present rational
maps which are good approximations to these Skyrmions and accurately
compute their energy by relaxation using the full non-linear dynamics.
\end{abstract}

\newpage
\section{Introduction}\news
\subsection{Overview}

The possibility that solitons can be used to represent particles is an
 attractive one, very much at the heart of current ideas in high
 energy physics. The first such model, known as the Skyrme
 model~\cite{Sk}, was proposed in 1961 as a theory for the strong
 interactions of pions. The resulting non-linear field theory admits
 topological soliton solutions, which became known as Skyrmions.  In
 the context of nuclear physics, the topological charge which
 stabilizes  these solitons was identified with baryon number and
 hence the solitons  themselves were identified with baryons. 

This model was set aside after the advent of gauge theories and
 Quantum Chromodynamics (QCD) in the late 1960's, but much
 later~\cite{WIT} it was shown to be a low-energy effective action for
 QCD, in the limit of the number of colours $(N_{\rm c})$ being
 large. Subsequent work has shown the model to be capable of describing
 at least some aspects of the low-energy behaviour of hadrons.  In
 particular, it was shown that the properties of the proton and
 neutron could be adequately described using the simple quantization
 of the solution with a single unit of topological charge~\cite{ANW},
 and that the other static 
 configurations appeared to be most stable when considering
 charges corresponding to $^{4}{\rm He}$ and $^{7}{\rm Li}$~\cite{BS2}.

Although understanding the properties of light nuclei is the main
motivation for this paper, we shall only comment very briefly at
various points in our discussion on the implications of our results
for this application. Instead, we will concentrate on the Skyrmions
themselves, which are of interest in their own right.  As examples of
three-dimensional topological solitons, they have been seen to be very
similar to BPS monopoles~\cite{AH,BS1,sutrev}.  From a mathematical
point of view they can be thought of as maps between
3-spheres and, therefore, the minimum energy configurations which we
compute are the minimum energy maps relative to the Skyrme energy
functional. Although there are no doubt many other possibilities for
an energy functional on the space of maps between 3-spheres, the
Skyrme functional is the simplest which supports topological
structures and, therefore, it is also interesting to speculate as to
their generality in this context.

In the following we will present an extensive study of minimum energy
solitons using two very different numerical methods\footnote{The bare
essentials of this work were first reported in ref.~\cite{BS4}. Here,
we give a more detailed exposition of our results.}.  With a small
number of caveats, the two approaches come to the same conclusion: that
there is a connection with fullerene cages~\cite{kroto} familiar in
carbon chemistry, as conjectured in ref.~\cite{BS2}, and that the
solutions can be represented in terms of the rational map
ansatz~\cite{HMS}. For each topological charge, we will discuss the
structure and symmetries of the solution in detail, and present an
explicit analytic formula for the approximate solution. In the small
number of cases where there is some deviation from the fullerene
hypothesis, we will discuss qualitatively,  and sometimes
quantitatively, possible
reasons. Finally, in the context of the two applications mentioned
above, nuclear physics and generalized harmonic 
maps between 3-spheres, we will attempt to
provide some heuristic insight into why our conclusions as to the
structure of Skyrmions are consistent with the particular application.

\subsection{Skyrmions}

In terms of the algebra valued  currents 
$R_{\mu}=(\partial_{\mu}U)U^{\dag}$ of an SU(2) valued field $U$, the
Skyrme Lagrangian density  is, 
\be
{\cal L} = \frac{1}{24\pi^2}\bigg[-{\rm Tr}\big{(}R_{\mu}R^{\mu}\big{)}
+{1\over 8} {\rm Tr}\big{(}[R_{\mu},R_{\nu}][R^{\mu},R^{\nu}]\big{)}\bigg]\,.
\label{lag}
\ee
At first sight the domain is $\R^3$ and the target space is the group
manifold of SU(2), $S^3$, but the finite energy boundary
condition, $U(\infty)=I$, means that $U$ is in fact a map from 
compactified $\R^3\sim S^3\mapsto S^3$. Such mappings have 
non-trivial homotopy classes characterized by $\pi_3(S^3)=\Z$, which
has the explicit representation
\be
B=- {1\over 24\pi^2}\epsilon_{ijk}\int d^3x
 {\rm Tr}\left(R_iR_jR_k\right)\,.
\label{defb}
\ee  A more geometrical description of the model can be made in terms
of the strain tensor defined at each point ${\bf x}$ in the domain by
\be 
 D_{ij}=-{1\over 2}{\rm Tr}\left(R_iR_j\right)\,, 
\label{strain}
\ee 
which can be
thought of as quantifying the deformation induced by the map between
3-spheres. This
symmetric, positive definite tensor can be diagonalized with
non-negative eigenvalues $\lambda_1^2$, $\lambda_2^2$, $\lambda_3^2$,
and the static energy, $E$, and baryon number, $B$, can be computed as
integrals over $\R^3$ of the corresponding densities ${\cal E}$ and
${\cal B}$ given by   
\be  {\cal
E}=\frac{1}{12\pi^2}(\lambda_1^2+\lambda_2^2+\lambda_3^2
+\lambda_1^2\lambda_2^2+\lambda_2^2\lambda_3^2+\lambda_3^2\lambda_1^2)\,,
\quad {\cal B}=\frac{1}{2\pi^2}\lambda_1\lambda_2\lambda_3\,.   
\ee 
A simple
manipulation of these expressions allows one to deduce the
Faddeev-Bogomolny bound  $E\ge \vert B\vert$, but in contrast to
monopoles and vortices this bound cannot be saturated for any
non-trivial finite energy configuration. In fact, it can be  attained
only when all the eigenvalues of the strain tensor are equal to one at all
points in space --- an isometry --- and this is clearly not possible
since $\R^3$ is not isometric to $S^3.$
 In practice we
shall see that minimization of the Skyrme energy functional associated
with the Lagrangian density requires that the solution be as close to
this bound as possible. The energy minimization can be
thought of as finding the map which is as close to an isometry as possible,
when averaged over space~\cite{krusch}.

The boundary condition breaks the chiral symmetry (${\rm SU}(2)\times
{\rm SU}(2)$) of the Skyrme model to an SO(3) isospin symmetry
$U\mapsto {\cal O}U{\cal O}^\dagger$, where ${\cal O}$ is a constant
element of SU(2). When we refer to a spatial  symmetry of a
Skyrmion, such as spherical symmetry, the fields are not invariant
under a spatial rotation, but rather there is an equivariance property
that the effect of a spatial rotation can be absorbed into an isospin
transformation. This implies that both the energy density, ${\cal E}$,
and baryon density, ${\cal B}$, are strictly invariant under the
symmetry.

The $B=1$ Skyrmion is spherically symmetric~\cite{Sk}, the maximally
allowed symmetry of the Lagrangian,  and its energy is
$E=1.232$~\cite{ANW}. In fact, spherically symmetric solutions exist
at all charges, but   they are not the minimum energy
configurations, which are less symmetric. The $B=2$ solution is axially
symmetric~\cite{KS,V} and the higher charge solutions all have point
symmetries~\cite{BTC,BS2} which are subgroups of O(3). For $B=3,4,7$
the Skyrmions have the Platonic symmetries of the tetrahedron ($T_d$),
the cube ($O_h$) and the dodecahedron ($Y_h$)
respectively, while for $B=5,6,8$ the Skyrmions
have the dihedral symmetries $D_{2d}$, $D_{4d}$ and $D_{6d}$ respectively
(see the discussion in section~\ref{sec-symmetric}
if you are unfamiliar with these point groups).
For $B=9$ a tetrahedrally symmetric Skyrmion has been found, but as we
shall discuss later this appears not to be the minimum energy
configuration, and hence is probably a low-energy saddle point.

In all the above cases the baryon density (and also the energy
density) is localized around the edges of a polyhedron.  From these
known results we were able, in a previous paper \cite{BS2}, to
formulate some simple geometrical rules\footnote{The original
Geometric Energy Minimization (GEM) rules were that all the solutions
computed for $B\le 9$ had the symmetries of almost spherical
trivalent polyhedra with $4(B-2)$ vertices, $2(B-1)$ faces and $6(B-2)$
edges.}
for the structure of Skyrmions
which led us to conjecture that higher charge Skyrmions ($B\ge 7$)
would resemble trivalent polyhedra formed from 12 pentagons and
$2B-14$ hexagons.  We will refer to such structures as fullerene-like
and to the conjecture as the fullerene hypothesis
since precisely the same structures arise in carbon chemistry where
carbon atoms sit at the vertices of such polyhedra, known as 
fullerenes~\cite{atlas}.  On the basis of this,  it was suggested that
the minimum energy Skyrmion of charge $B$, ${\cal S}_{B}$, would have
the same symmetry as a fullerene from the family ${\rm
C}_{4(B-2)}$. For low charges ($B=7$, $B=8$) this leads to a unique
prediction for ${\cal S}_B$, and indeed this was what we found in our
original simulations. But as the charge increases the number of
possible structures increases, in particular for $B=9$ there are 2
possibilities with $D_2$ and $T_d$ symmetries respectively, for $B=10$
there are 6, for $B=11$ there are 15, with a rapid increase for
$B>11$. Using this simple analogy, it was possible to predict that
there would be an icosahedral configuration with $B=17$ corresponding
to the famous Buckminsterfullerene structure of ${\rm C}_{60}$, and given
its highly symmetric structure we suggested that this would be the
minimum energy configuration at that charge.

Our original work on computing minimum energy Skyrmions~\cite{BS2} and
also that of ref.~\cite{BTC} used the full non-linear field equations,
or the full non-linear energy functional; a very resource hungry
procedure even in the modern era of parallel supercomputers. It is
also likely to be somewhat imprecise since (1) it can be very difficult to
identify the particular symmetries of the solution that one computes
in this way, even using sophisticated visualization
packages, and (2) the choice of initial conditions is somewhat
arbitrary: even when they are chosen to have no particular
symmetry, it is impossible to guarantee that they will relax to the
global minimum. Fortunately, help is at hand in the form of the rational map
ansatz~\cite{HMS}, which was devised as an approximate representation
of the solutions computed in ref.~\cite{BS2}. Although this
representation is not exact it reduces the degrees of freedom in the
problem to be a finite, manageable number, allowing computations to
take place in an acceptable amount of time. Of course, the original
approach based on the full non-linear equations still has a role to
play in firstly checking and then fully relaxing the approximate
solutions computed in the rational map approach. The development of
this subject now treads an interesting interface between numerically
generated solutions and this analytic approximation.

In this paper we shall first compute what are the minimum
energy solutions of the Skyrme model upto $B=22$, under the assumption
that they can be adequately represented by the rational map
ansatz. Then by simulating collisions of well separated Skyrmion
clusters using the full non-linear field equations, followed by a
numerical relaxation of the resulting coalesced cluster, we shall
attempt to verify that these structures are in fact the minimum energy
Skyrmions.  We find precisely the fullerene structures conjectured in
ref.~\cite{BS2} for all but a small number of cases where there are
interesting caveats.  We also demonstrate that not only are the
minimal energy Skyrmions fullerene-like, but that there are also
several other fullerene Skyrmions at a given baryon number whose
energies are only very slightly higher than the minimal
value. Finally, using the full non-linear equations once again, we
relax the rational map generated solutions with specified symmetries
to compute accurately the energies of the configurations.

\subsection{Rational map ansatz}
\label{sec-rma}

The rational map ansatz was introduced in ref.~\cite{HMS}, and is a
way to construct approximate Skyrmions from rational maps between
Riemann spheres. Briefly, we use spherical coordinates in $\R^3$, so
that a point ${\bf x}\in\R^3$ is given  by a pair $(r,z)$, where
$r=\vert{\bf x}\vert$ is the  distance from the origin, and $z$ is a
Riemann sphere coordinate giving the point on the unit two-sphere
which intersects the half-line through the origin and the point ${\bf
x}$.

Now, let $R(z)$ be a degree $B$ rational map between Riemann spheres,
that is, $R=p/q$ where $p$ and $q$ are polynomials in $z$ such that
$\max[\mbox{deg}(p),\mbox{deg}(q)]=B$,  and $p$ and $q$ have no common
factors.  Given such a rational map the ansatz for the Skyrme field is
\be  U(r,z)=\exp\bigg[\frac{if(r)}{1+\vert R\vert^2} \pmatrix{1-\vert
R\vert^2& 2\bar R\cr 2R & \vert R\vert^2-1\cr}\bigg]\,,
\label{rma}
\ee where $f(r)$ is a real profile function satisfying the  boundary
conditions $f(0)=\pi$ and $f(\infty)=0$, which is determined by
minimization of the Skyrme energy of the field (\ref{rma}) given a
particular rational map $R$.  It can be shown that this Skyrme field
has charge $B$, and for $1\le B\le 9$ rational maps were presented in
ref.~\cite{HMS} which reproduced Skyrmions with the same
symmetries as those computed in ref.~\cite{BS2}.  Furthermore, they
were shown to  have energies which are only about one or two percent
above the numerically calculated values.

Substitution of the rational map ansatz (\ref{rma}) into the Skyrme
energy functional results in the following expression for the energy \be
E=\frac{1}{3\pi}\int \bigg( r^2f'^2+2B(f'^2+1)\sin^2 f+\I\frac{\sin^4
f}{r^2}\bigg) \ dr\,,
\label{rmaenergy}
\ee where $\I$ denotes the integral \be \I=\frac{1}{4\pi}\int \bigg(
\frac{1+\vert z\vert^2}{1+\vert R\vert^2}
\bigg\vert\frac{dR}{dz}\bigg\vert\bigg)^4 \frac{2i \  dz  d\bar z
}{(1+\vert z\vert^2)^2}\,.
\label{i}
\ee To minimize the energy (\ref{rmaenergy}), therefore,  one first
determines the rational map which minimizes $\I$, which may be thought
of as an energy functional on the space of rational maps. Then given
the minimum value of $\I$ it is a simple exercise to find the profile
function which minimizes the energy (\ref{rmaenergy}) using a
gradient flow method to solve the appropriate boundary value
problem. Thus, within the rational map ansatz, the problem of finding
the minimal energy Skyrmion reduces to the simpler problem of
calculating the rational map which minimizes the function
$\I$. Computing the map which minimizes this set up is the
essence of our procedure for finding the minimal energy Skyrmion, and
in section \ref{sec-numerical} we shall describe our numerical
techniques used to address this problem.

The baryon density is proportional to the derivative of the rational
map, and (counting multiplicities) this will have
$2B-2$ zeros, giving the points on the Riemann sphere for
which the baryon density vanishes along the corresponding half-lines
through the origin. In terms of a baryon density isosurface plot these
correspond to holes in a shell-like structure which 
resembles a polyhedron and the holes correspond to the face centres.

\subsection{Symmetric maps : general discussion}
\label{sec-gensymm}

Since the maps we shall be dealing with describe symmetric Skyrmions,
let us recall what it means for a rational map (and hence the
associated Skyrmion) to be symmetric under a group $G\subset SO(3).$
Consider a spatial rotation $g\in SO(3)$, which acts on the Riemann
sphere coordinate $z$ as an $SU(2)$ M\"obius transformation \be
z\mapsto g(z)=\frac{\gamma z +\delta}{-\bar\delta z+\bar\gamma}\,,
\qquad \mbox{where} \qquad \vert \gamma\vert^2+\vert \delta\vert^2=1.
\label{mob1}\ee
Similarly a rotation, $D\in SO(3)$, of the target two-sphere (which
corresponds to an isospin transformation) will act in the same way \be
R\mapsto D(R)=\frac{\Gamma R +\Delta}{-\bar\Delta R+\bar\Gamma}\,,
\qquad\mbox{where} \qquad \vert \Gamma\vert^2+\vert \Delta\vert^2=1.
\label{mob2}\ee
A map is $G$-symmetric if, for each  $g\in G$, there exists a target
space rotation, $D$, which counteracts the effect of the spatial
rotation, that is, \be  R(g(z))=D(R(z)). 
\label{defsym}
\ee Note that in general the rotations on the domain and target
spheres will not be the same, so that $(\gamma,\delta)\ne
(\Gamma,\Delta).$ 

Since we are dealing with $SU(2)$ transformations the set of target
space rotations will form a representation of the double group of $G$,
which is the group of order $2\vert G\vert$ obtained from $G$ by the
addition of  an element $\bar E$ which squares to the identity.  The
fact that we are dealing with the double group is important since it
has representations which are not representations of $G.$ From now on
it is to be understood that when we refer to a group $G$ we shall
actually mean its double group.

To determine the existence and compute particular symmetric rational
maps is, therefore, a matter of classical group theory. We are
concerned with degree $B$ polynomials which form the carrier space for
$\underline{B+1}$, the $(B+1)$-dimensional  irreducible representation
of $SU(2).$ Now, as a representation of $SU(2)$ this is irreducible,
but if we only consider the restriction to a  subgroup $G$,
$\underline{B+1}\vert_G$, this will in general be reducible. What we
are interested in is the irreducible decomposition of this
representation and tables of these subductions  can be found, for
example, in ref.~\cite{pgtt}.

The simplest case in which a $G$-symmetric degree $B$ rational map
exists is if  \be \underline{B+1}\vert_G=E+...  \ee where $E$ denotes
a two-dimensional representation. In this case a basis for $E$
consists of two degree $B$ polynomials which can be taken to be the
numerator and denominator of the rational map.  A subtle point which
needs to be addressed is that the two basis polynomials may have a
common root, in which case the resulting rational map is degenerate
and does not correspond to a genuine degree $B$ map.

More complicated situations can arise, for example, if \be
\underline{B+1}\vert_G=A_1+A_2+...  \ee  where $A_1$ and $A_2$ denote
two one-dimensional representations, then a whole one-parameter family
of maps can be obtained by taking a constant multiple of the ratio of
the two polynomials which are a basis for $A_1$ and $A_2$
respectively.  An $m$-parameter family of $G$-symmetric maps can be
constructed if the decomposition contains $(m+1)$ copies of a
two-dimensional representation, that is, \be
\underline{B+1}\vert_G=(m+1)E+... \ee where the $m$ (complex)
parameters correspond to the freedom in the decomposition of $(m+1)E$
into $m+1$ copies of $E.$

Explicit examples corresponding to the above types of decompositions
will be constructed in section~\ref{sec-symmetric}.  For a detailed
explanation of how to calculate these maps by computing appropriate
projectors see ref.~\cite{HMS}.

\subsection{Symmetric maps : specific examples}
\label{sec-symmetric}

As we shall discuss in section~\ref{sec-simanneal} the basic procedure
for finding the minimum energy Skyrmion in the rational map ansatz
will be to minimize the function $\I$ subject to the map 
having degree $B$. However, this will find the
map in an arbitrary spatial orientation, preventing identification
of the symmetry directly from the map. One can always compute the
corresponding baryon density isosurface and identify the symmetry by
eye, a procedure which is often helpful,
but this is also fraught with difficulties, particularly when,
for example, the solution only has a small number of symmetry
generators. Therefore, in order to be sure of the symmetry
identification we will also search maps which are restricted to  have a
particular symmetry. 

Since all the point groups that we shall consider
are subgroups of O(3), they must be either
cyclic groups, $C_n$, which involve invariance under rotations by
$(360/n)^{\circ}$ about some axis, dihedral groups, $D_n$, which are
obtained from the cyclic group by the addition of a $C_2$ axis which
is perpendicular to the main symmetry axis, tetrahedral groups ($T$),
which are the  symmetries associated with the tetrahedron, octahedral
groups ($O$), those associated with the octahedron/cube, or
icosahedral groups ($Y$), those associated with the
icosahedron/dodecahedron. Each of these symmetry groups can be
extended by the inclusion of reflections. All the icosahedral maps
presented in this paper have already been discussed in
ref.~\cite{HMS}, while all the octahedral maps used
are easily deduced from the tetrahedral
maps discussed below, so we shall only concentrate on understanding
the details of the dihedral and tetrahedral maps.

In terms of the Riemann sphere coordinate $z$ the generators of the
dihedral group $D_n$ may be taken to be $z\mapsto e^{2\pi i/n}z$ and
$z\mapsto 1/z.$ This can be extended by the addition of a reflection
symmetry in two ways: by including a reflection in the plane
perpendicular to the main $C_n$ axis, which is represented on the
Riemann sphere by invariance under  $z \mapsto
1/\bar z$, and the group $D_{nh}$ is obtained.  Alternatively, a
reflection symmetry may be imposed in a plane which contains the main
symmetry axis and bisects the $C_2$ axes obtained by applying the
$C_n$ symmetry to the $C_2$ axis. This reflection is represented on
the Riemann sphere as invariance under $z \mapsto  e^{\pi i/n} \bar
z$, and the resulting group is $D_{nd}.$

To construct $D_n$ symmetric maps does not require any group theory 
formalism discussed in section~\ref{sec-gensymm}
since it is a simple task to explicitly apply the two generators
of $D_n$ to a general degree $B$ rational map to determine a family of
symmetric maps. Explicitly, an $s$-parameter family is given
by\footnote{There are other $D_{n}$ symmetric families of maps in
addition to those of the form (\ref{dnd}), but they will not be needed in this
paper.}
\be
R(z)=\bigg(\sum_{j=0}^s a_jz^{jn+r}\bigg)/\bigg(\sum_{j=0}^s
a_{s-j}z^{jn}\bigg)\,,
\label{dnd}
\ee
where $r=B\ \mbox{mod\ } n$ and $s=(B-r)/n.$
Here $a_s=1$ and $a_0,..,a_{s-1}$ are arbitrary complex parameters.
Clearly, this map satisfies the conditions for it to be symmetric
under $D_n$,
\be
R(e^{2\pi i/n}z)=e^{2\pi ir/n}R(z)\,,\,R(1/z)=1/R(z)\,,
\ee
and imposing a reflection symmetry constrains the otherwise  
complex coefficients $a_j$ to either be real, or pure imaginary.
In the case of $D_{nh}$ symmetry the
condition is that all $a_j$ are real, whereas for $D_{nd}$ symmetry
the coefficient $a_j$ is either real or purely imaginary depending
on whether $(s-j)\ \mbox{mod 2\ }$ is zero or one respectively

The procedure for constructing tetrahedrally symmetric maps is more
difficult than for dihedral symmetries and the systematic group theory approach
discussed in section~\ref{sec-gensymm} must be employed. No simple
formula such as (\ref{dnd}) exists, so for later reference we shall,
therefore, need to recall the basic facts about the irreducible representations
of the tetrahedral group $T.$

$T$ has three one-dimensional representations, which are the trivial
representation, $A$, and two conjugate representations $A_1$ and $A_2.$
There is also a three-dimensional representaion, $F$, which is obtained
as $\underline{3}\vert_T.$ In addition to these representations there
are three two-dimensional representations of the double group of $T$,
which we denote by $E',E_1',E_2'$, where the prime signifies that these
are not representations of $T$, but only of the double group of $T.$
$E'$ is obtained as $\underline{2}\vert_T$ and $E_1'$ and $E_2'$ are
conjugate representations. 

\section{Numerical minimization algorithms}\news
\label{sec-numerical}

\subsection{Overview}

Minimization of an energy functional is a classical numerical problem,
with no hard and fast optimum method. Methods which are tailored for a
particular application can work very  badly in others. In this section
we will outline the basic features of the numerical methods which we
have employed to compute minimum energy Skyrmions with and without
using the rational map ansatz. We will attempt to
discuss both the advantages and disadvantages of the two methods.

Our original approach to this problem was to use the code first used in
ref.~\cite{BS1} to evolve the full field equations for well-separated
Skyrmions, and this is discussed in section~\ref{sec-fulldyn}. It
worked well for $B<9$~\cite{BS2} and its results were the main
motivation for the rational map ansatz. Its disadvantages are that it
is  very slow, requiring many hours of CPU time on a parallel
computer, and in circumstances where there is the possibility of  two
or more minima  separated by a small energy gap, dependence on the
choice of initial conditions is also an issue. It is, however, the
only way in which the results of approximate methods, such as the
rational map ansatz, can be checked. Creating a particular
configuration from many different initial conditions using this method
can be thought of as strong evidence for it being the minimum energy
solution, irrespective of other considerations.

The simulated annealing of rational maps as discussed in
section~\ref{sec-simanneal} is by contrast fast --- it only requires a
serial processor --- by virtue of the fact that the number of degrees of
 freedom have been substantially reduced, is relatively independent of the
initial conditions and much less sensitive to local minima. But, of course,
the results are only as good as the rational map approach to
describing Skyrmions. Thankfully, as we shall see, the rational map
approach in general works very well, allowing us to generate symmetric
Skyrmions with large baryon number.

\subsection{Simulated Annealing of Rational Maps}
\label{sec-simanneal}

Simulated annealing is a fairly recent numerical method for obtaining
the global minimum of an energy function, and is based
on the way that a solid cools to form a lattice \cite{sabook}.
Let $E(\bf{a})$ be an energy function which depends on a number of parameters
$\bf{a}.$ The idea is that at a given temperature, $T$, the system
is allowed to reach thermal equilibrium, characterized by the
probability of being in a state with energy ${\cal E}$ given by the
Boltzmann distribution
\be
\mbox{Pr}(E={\cal E})=\frac{1}{Z(T)}\exp(-{\cal E}/T)\,,
\ee
where $Z(T)$ is the partition function.

In practice, this is achieved by applying
a Metropolis method:
starting with a given configuration, $\bf{a}$, a small random 
perturbation $\bf{\delta a}$ is made and the energy of the resulting
configuration is computed. If the change in energy,
$\delta E=E({\bf{a+\delta a}})-E(\bf{a})$, is negative then the
new configuration is accepted, that is, ${\bf a}$ is replaced by
${\bf a+\delta a}.$ However, if the change results in an increase in
energy then the probability of accepting the new configuration is
$e^{-\delta E/T}.$ By performing a large number of such perturbations
thermal equilibrium can be achieved at temperature $T$.

The procedure to minimize the energy is to start at a high
temperature, bring the system into thermal equilibrium and then  lower the
temperature before regaining the equilibrium.
As the temperature is decreased, the system is more
likely to be found in a state with lower energy and in the limit as
$T\rightarrow 0$ the configuration will move toward a minimum of $E.$
In the limit of infinitesimally slow variations in the temperature
this can be shown to be the global minimum.

From the above description it is immediately clear that the  simulated
annealing method has a major advantage over other conventional
minimization techniques in that changes which increase the energy are
allowed, enabling the algorithm  to escape from
minima that are not the global minimum.  Of course, in practice one is
not guaranteed to find the global minimum since the number of
iteration loops used to bring the system into thermal equilibrium at a
fixed temperature and also the number of times the temperature can be
decreased are both restricted by computational resources. However, it
does provide the most efficient means for searching for a global
minimum and with sufficient computational resources, plus sufficient
care in applying them, one can be fairly confident of the final result. 

For our application to rational maps we obviously take the  energy
function to be $\ci$ and the parameters ${\bf a}$ to be the constants
in the rational map. To compute $\ci$ involves a numerical integration
over the sphere, which can be performed with standard methods, and in
a typical simulation this needs to be calculated approximately a
million times for a full simulated annealing run. In each case we take
the initial rational map to be the axially symmetric one $R=z^B,$
which for large $B$ has a very high value of $\I.$ 

Note that since we are using the Riemann sphere coordinate $z$, the
two-sphere metric is the Fubini-Study metric in this coordinate
system. This means that in terms of moving energy around the sphere
there is a bias between points near the south pole as compared to
those near the north pole, in terms of small variations of the
rational map parameters.  To counteract this discrepancy we perform a
spatial rotation of the configuration, $z\mapsto 1/z$ each time the
temperature is decreased.

To identify the rational map, and more importantly  its symmetries,
 produced by the simulated annealing algorithm is a two stage
 process. First, we compute the minimum energy map assuming no
 particular symmetry, that is, we allow a general map of degree
 $B$. As already pointed in section~\ref{sec-symmetric} the end result
 will be a rational map which is in a random orientation in both the
 domain and target two-spheres.  The target space orientation could
 easily be fixed (in fact, it is more convenient not to do this, due to
 the above comments regarding  a periodic spatial rotation of the map
 during the minimization procedure), but there is no simple way to
 make the spatial orientation such that the symmetry generators of the
 map are conveniently represented.  Once one has the minimizing degree
 $B$ rational map and  the corresponding minimum value of $\ci,$ the
 Skyrmion is then constructed and its baryon density is plotted and
 examined in an attempt to identify its symmetries by eye. This conjectured
 symmetry is then confirmed by constructing the most general map with
 this symmetry and minimizing within this constrained symmetric family
 to check that the same minimum value of $\ci$ is recovered. As a
 final check the corresponding Skyrmion is constructed and its baryon
 density examined to confirm that it is identical to the one obtained
 previously. 

For each charge we have also performed several simulated annealing
 runs within constrained symmetric families, such as $D_2$, $D_3$ and
 $D_4$. Since all the symmetry groups must be subgroups of O(3), the number of
 possibilities is finite and checking just these three possibilities,
 allows one to rule out a large fraction of them; the rest often being
 possible by eye. This not only provides an additional check that the
 minimizing map was found, but also allows us to obtain other low
 energy maps, which may be either saddle points or local minima. 

Finally, it is perhaps worth mentioning that a simulated annealing method
has recently been used in a rather different way to study Skyrmions \cite{HSW}.
These authors used a simulated annealing algorithm on a discretized 
version of the full Skyrme energy, taking the parameters to be the field
values at the discretized lattice sites. This is a much more computationally
expensive approach than using the rational map ansatz, which is 
probably the reason these authors only considered Skyrmions upto
charge $B=4$, where, of course, the results were already known.
Nonetheless, it is interesting to know that a simulated annealing
approach is viable in this manner and demonstrates yet another 
application of this versatile technique.

\subsection{Full Field Dynamics} 
\label{sec-fulldyn}
\subsubsection{Sigma Model Formulation}
In the $SU(2)$ form, the equations of motion are cumbersome to handle numerically, so we
 convert to the notation of a non-linear sigma model (NLSM), which has Lagrangian,
\be
{\cal L} =\partial_{\mu}{\phi}\cdot\partial^{\mu}{\phi}-\textstyle{1\over 2}
\big{(}\partial_{\mu}{\phi}\cdot\partial^{\mu}{\phi}\big{)}^2 +
\textstyle{1\over 2}\big{(}\partial_{\mu}{\phi}\cdot\partial_{\nu}
{\phi}\big{)}\big{(}\partial^{\mu}{\phi}\cdot\partial^{\nu}{\phi}\big{)}
+\lambda(\phi\cdot\phi-1)\,,
\ee
with the Lagrange multiplier $\lambda$ introduced to maintain 
the constraint $\phi\cdot\phi=1$.

The Euler-Lagrange equations are given by 
\be
\big{(}1-\partial_{\mu}{\phi}\cdot\partial^{\mu}{\phi}\big{)}
\square{\phi}-\big{(}\partial^{\nu}{\phi}\cdot\partial_{\mu}\partial_{\nu}
{\phi}-\partial_{\mu}{\phi}\cdot\square{\phi}\big{)}\partial^{\mu}
{\phi}+\big{(}\partial^{\mu}{\phi}\cdot\partial^{\nu}{\phi}\big{)}
\partial_{\mu}\partial_{\nu}{\phi}-\lambda{\phi} =0\,,
\label{eleqn}
\ee
where the Lagrange multiplier can be calculated by contracting 
(\ref{eleqn}) with $\phi$ and using the second derivative of the constraint,
\bea
\lambda=&\big{(}1-\partial_{\mu}{\phi}\cdot\partial^{\mu}
{\phi}\big{)}{\phi}\cdot\square{\phi} + \big{(}\partial^{\mu}
{\phi}\cdot\partial^{\nu}{\phi}\big{)}\big{(}{\phi}\cdot
\partial_{\mu}\partial_{\nu}{\phi}\big{)} &\cr 
= &-\big{(}\partial_{\mu}{\phi}\cdot\partial_{\nu}{\phi}\big{)}
\big{(}\partial^{\mu}{\phi}\cdot\partial^{\nu}{\phi}\big{)}
-(1-\partial_{\mu}{\phi}\cdot\partial^{\mu}{\phi}\big{)}
\partial_{\nu}{\phi}\cdot\partial^{\nu}{\phi}. &
\label{lambd}
\eea
Denoting differentiation with respect to time as a dot, these equations can be recast as 
\be
M\ddot{\phi}-{\alpha}\big{(}\dot{\phi},\partial_i{\phi},
\partial_i\dot{\phi},\partial_i\partial_j{\phi}\big{)}-\lambda{\phi}=0\,,
\ee
where the symmetric matrix $M$ has elements 
\be
M_{ab}=\big{(}1+\partial_j{\phi}\cdot\partial_j{\phi}\big{)}
\delta_{ab}-\partial_j\phi_a\partial_j\phi_b\,,
\ee
and $\alpha$ is given by 
\bea
&{\alpha}= \big{(}\dot{\phi}\cdot\partial_i\partial_i{\phi}-\partial_i{\phi}\cdot\partial_i\dot{\phi}\big{)}\dot{\phi}
+2\big{(}\dot{\phi}\cdot\partial_i{\phi}\big{)}\partial_i
\dot{\phi}-\big{(}\dot{\phi}\cdot\partial_i\dot{\phi}\big{)}
\partial_i{\phi} 
-\dot{\phi}^2\partial_i\partial_i{\phi} \cr
&+\big{(}\partial_i{\phi}\cdot\partial_i\partial_j{\phi}
-\partial_j{\phi}\cdot\partial_i\partial_i{\phi}\big{)}
\partial_j{\phi}+\big{(}1+\partial_j{\phi}\cdot\partial_j{\phi}\big{)}\partial_i\partial_i{\phi}-\big{(}\partial_i{\phi}\cdot
\partial_j{\phi}\big{)}\partial_i\partial_j{\phi}.& \cr
&      
\eea
Quite clearly these equations of motion are not analytically tractable. 
In subsequent sections we will discuss our numerical methods for evolving
 the equations of motion for spatially discretized 
initial conditions and for obtaining minimal energy static Skyrmions.
 As we shall see this is still a highly non-trivial task
 and can only be done for a specialized, but ill-defined, set of initial 
conditions.

\subsubsection{Discretization and boundary conditions}

There are three aspects common to almost all numerical approaches to solving
 non-linear PDE's. The first is a spatial discretization and an approximation
 for spatial derivatives. We discretized on a regular, cubic grid with
 $N$ points in each of the Cartesian directions and the array 
$\phi_{i,j,k}\approx\phi(i\Delta x, j\Delta x, k\Delta x)$. The choice of
 $N$ and $\Delta x$ is of critical importance, since the soliton 
configurations we wish to represent are localized. We found that grids 
with $N=100$ and $\Delta x=0.1$ were convenient for representing all the
 configurations which we study in this paper, although larger grid spacings
 $(\Delta x=0.2)$ also give sensible results, and larger grids ($N=200$)
were used to obtain more accurate calculations of energies.

The spatial derivatives used were fourth order, so as to accurately represent
 the large spatial gradients of the solitonic configurations. Since the 
reader may not be totally familiar with this procedure, various 
expressions for derivatives are presented below: for first order derivatives,
\be
{\partial\phi\over\partial x}={-\phi_{i+2,j,k}+8\phi_{i+1,j,k}-8\phi_{i-1,j,k}+
\phi_{i-2,j,k}\over 12\Delta x}+{\cal O}(\Delta x^4)\,,
\ee
for second order derivatives,
\be
{\partial^2\phi\over\partial x^2}={-\phi_{i+2,j,k}+16\phi_{i+1,j,k}-
30\phi_{i,j,k}+16\phi_{i-1,j,k}-\phi_{i-2,j,k}\over 12\Delta x^2} +{\cal O}(\Delta x^4)\,,
\ee 
and for mixed second order derivatives
\bea
&\displaystyle{
\partial^2\phi\over \partial x^2}
+2{\partial^2\phi\over\partial x\partial y}
+{\partial^2\phi\over\partial y^2}={\cal O}(\Delta x^4)+ \\
&({-\phi_{i+2,j+2,k}+16\phi_{i+1,j+1,k}-30\phi_{i,j,k}+16\phi_{i-1,j-1,k}
-\phi_{i-2,j-2,k})/( 12\Delta x^2)}\,. \nonumber
\eea

The next part of the procedure is a method for time evolution. 
The equations of motion can be transformed into first order form,  
\be 
M\dot\psi - \alpha(\psi,\partial_i\phi,\partial_i\psi,
\partial_i\partial_j\phi)-\lambda\phi=0\,,
\ee
by defining $\psi=\dot \phi$, and this can be solved using a leapfrog method.
 This involves replacing 
\be 
\dot\phi= {\phi^{+}-\phi^{-}\over2\Delta t} + {\cal O}(\Delta t^2)\,,\quad
\dot\psi= {\psi^{+}-\psi^{-}\over2\Delta t} + {\cal O}(\Delta t^2)\,,
\ee
where $+$ and $-$ correspond to the values of $\psi$ and $\phi$ at one 
step after and one before respectively. In a much simpler case such as the
 wave equation, this creates a decoupling of the arrays containing the 
discretized versions $\phi$ and $\psi$. However, the non-linear dependence 
of the function $\alpha$ on $\psi$ requires the storing of two copies of
 each array, and hence four in total, requiring 64Mb of
 core memory for $N=100$.
 The choice of $\Delta t$ is also crucial, since it can 
create numerical instability. The standard Courant condition for a linear 
equation in three dimensions states that $\sqrt{3}\Delta t< \Delta x$.
 The non-linear nature of the Skyrme equations leads to a non-trivial
 modification to this relation, which still stands as the best possible due 
to reasons of causality. We find, essentially by trial and error, that
 $\Delta t \approx\Delta x/10$ leads to a stable algorithm. We should
 note at this stage that there is another, potentially more pathological, instability of the Skyrme model 
which is discussed in a subsequent section.

Finally, one is required to specify boundary conditions for the finite
grid employed. Since the fourth order spatial derivatives require a
five  point wide stencil, one point away from the boundary it is
necessary to  use second order spatial approximations. But on the
boundary itself the spatial derivatives cannot be evaluated, since the
second order spatial approximation  requires a three point wide
stencil. We experimented with various different types of boundary
conditions, such as Neumann (zero normal derivative), Dirichlet
(fixed) and periodic. The results presented here are for Dirichlet
boundary conditions, although we believe that the use of  Neumann
boundaries would have little effect on the results.

\subsubsection{Imposing the constraint}

In the previous section we have discussed all aspects of the numerical
 solution of the Skyrme equations of motion which are common to the
 numerical solution of most non-linear PDE's. There is, however, an
 added extra which makes life much more difficult in the case of a
 NLSM. In the previous section we included the Lagrange multiplier
 $\lambda$, assuming that it can be calculated from $\psi$, $\phi$
 and their spatial derivatives.  In fact, this leads to a numerical
 scheme which becomes unstable for any choice of $\Delta t$ in under
 ten timesteps. The problem is that  we have ignored the reason for
 its introduction; to maintain the constraint $\phi\cdot\phi=1$, which
 is manifest in the NLSM. A number of approaches have been developed
 to deal with such constraints.

Firstly, one could modify the numerical scheme to  calculate $\lambda$
so that it explicitly maintains the constraint for the discretized
equations of motion. At each step, this  is seen to be almost equal to
calculating $\lambda$ from the  formula (\ref{lambd}), but the two
differ by numerical discretization  effects at a level well below 1\%,
which nonetheless cause the solution to slip off the unit sphere, if
allowed to accumulate.

An alternative approach, which was found to work well in simulations
of Baby Skyr\-mions~\cite{wojtek}, is to simply rescale the field to
have unit modulus, that is, to continually make the replacement \be
\phi\mapsto {\phi\over\sqrt{\phi\cdot\phi}}\,,
\label{rescale}
\ee    at each point on the discretized grid after each timestep.
While appearing ugly from a purist  numerical analysis point of view,
this technique is effective and does not become unstable except in the
most  extreme circumstances. One is effectively projecting the field
back onto the sphere along the field itself, which has no particular
physical motivation, but if the modification is small, which can be
arranged by choosing a sufficiently small timestep, then this should
be as good as any other arbitrary choice.

Another possibility is to require that the derivative of the
constraint is zero: the relation $\phi\cdot\phi=1$ not only implies
that the solution lies on the unit sphere, but that it cannot come
off, that is, all the derivatives of the constraint are also
satisfied. This leads to an infinite hierarchy of relations which must
hold.  Obviously, with a second order time evolution one cannot hope
to maintain them all explicitly in the discretized system. However, if
one manages to satisfy the first one, it may be possible to construct
an effective code. It is possible to impose that $\phi\cdot\psi=0$ by
computing $\lambda$ in order to satisfy $\phi^{+}\cdot\psi^{+}=0$
which requires that, 
\be \lambda=-{\psi^-\cdot\psi +\phi\cdot
M^{-1}\alpha+ 2\psi\cdot M^{-1}\alpha\Delta t\over \phi M^{-1}\phi
+2\psi\cdot M^{-1}\phi\Delta t}\,.  
\ee
Although we should note that
this does not implicitly imply the imposition of the constraint on the
discretized equations.

Largely by trial and error, we find that the best method for our
numerical scheme is a hybrid of the last two methods.  It just so
happens that this is possible within our scheme, since there are two
discretized grids with essentially their own time evolution. Firstly,
we calculate $\lambda$ to satisfy $\phi^{+}\cdot\psi^{+}=0$, followed
by the rescaling transformation (\ref{rescale}) on the field
$\phi^{+}$.  We find that this hybrid methods maintains the constraint
for many thousands of timesteps.

\subsubsection{Non-Hyperbolic regions}

We have already discussed the potential for our numerical scheme to
 become  unstable  because of Courant-type instability and also due to
 the imperfect imposition of the constraint $\phi\cdot\phi
 =1$. However, there is a much more pathological instability  which
 comes about since the equations of motion are not manifestly
 hyperbolic and their  precise nature, hyperbolic, parabolic or
 elliptic, depends on the particular configuration being evolved
 \cite{CB}.

The problem is that the specific numerical scheme we have designed
 will work only for the hyperbolic case and we know of no way of
 treating all configurations within a single numerical
 scheme. Fortunately, the hyperbolic regime is the only physically
 meaningful one. One can  understand this by thinking of the Skyrme
 model as a low energy effective action, with higher order terms
 ignored. At higher energies, where the terms which are ignored would
 be large, the Skyrme equations of motion become non-hyperbolic.

It was suggested in ref.\cite{CB} that the equations of motion became
 non-hyperbolic whenever the kinetic energy is greater than the
 potential energy. Empirically, we find  that this is at least
 partially true, with an instability associated with a very large
 kinetic energy density relative to that of potential energy. However,
 this statement is  a little imprecise since we found that sometimes
 the kinetic energy density rose above the potential locally  on the
 discretized grid without creating instability.  Unfortunately, the
 complicated nature of the equations of motion makes it almost
 impossible to say for certain, which configurations will eventually
 lead to an instability and which will not, although experience has
 taught us that most of the configurations we wish to evolve for
 physical applications, such as  low energy nuclear physics, are
 possible.

\subsubsection{Locating minima}
\label{prop-locate}

In the preceding sections we have discussed how to construct a
numerical scheme  which  evolves the full non-linear equations of
motion for the Skyrme model. This allowed us  to simulate the dynamics
of Skyrmion collisions in ref.~\cite{BS1}. However, for the purposes
of this paper one would also like to create static multi-soliton
configurations. We are assisted in this  by the observation that when
well separated Skyrmions coalesce they seem to create low energy
symmetric multi-soliton states. 

The procedure that we use is to set up initial conditions with the
required topological  charge  involving a collision between one or
more Skyrmions of some particular charge, which can be done using the
rational map ansatz, described in section \ref{sec-rma}, using charges
for which the rational map is already known.  We then evolve the
configuration until they visibly coalesce or until the potential
energy begins to increase. At this point all the kinetic energy is
removed, that is, $\dot\phi=0$ at all points on the grid, and the
evolution is continued until once again the potential energy
rises. This procedure is repeated many times until the energy is no
longer decreasing. Once the solution is sufficiently relaxed it often
pays to just evolve the solution under the full equations of motion
without removing kinetic energy since this can prevent the
minute oscillations required to achieve the minimum from gaining
momentum.

Even this procedure  can be very slow, but can be speeded up by adding
a dissipative term to the  equations of motion, \be
M\ddot{\phi}-{\alpha}\big{(}\dot{\phi},\partial_i{\phi},
\partial_i\dot{\phi},\partial_i\partial_j{\phi}\big{)}-\lambda{\phi}=
-\epsilon\dot\phi\,, \ee for $\epsilon>0$. Clearly, the static
solutions are still the same. However, the dissipation causes the
solution to roll down the potential well to the minimum much quicker
in certain circumstances, with $\epsilon=0.5$ seeming to work well. We
should note that the addition of  this dissipation can also effect the
Courant instability of the  algorithm, and, in particular, very near
to the minimum one is plagued by instability.  Experience, has shown
us that a combination of running with dissipation and then without
helps speed up the process.

Obviously, one should be concerned that this process might not
 necessarily lead one to the global minima. One might, for example,
 relax down to a metastable local minima or the initial conditions may
 have some symmetry which is maintained by the equations of motion and
 hence the final solution. We attempt to ensure that we do not
 encounter the latter possibility by  creating initial conditions
 using the product ansatz, which is manifestly asymmetric  $(U_1U_2\ne
 U_2U_1)$. The possibility of local minima, however, can never be
 totally excluded, but one can build up confidence in the minima by
 using different initial configurations.

For low charges ($B\le 4$), the attractive channel configurations
discussed in ref.\cite{BS1} are particularly good initial
conditions. But for higher charge no such maximally attractive
channels exist for $B$ well separated Skyrmions and only a small
number of attractive configurations are known. We find that sensible
initial conditions can be produced for any charge $B>10$ by using two
clusters, one with charge $B-n$ and the other with charge $n$, such
that $n=1,2,3,4,5$ but no larger. These are Lorentz boosted together
with a velocity of $v=0.3$ in a collision which has a small but
non-zero impact parameter.

\begin{table}
\centering
\begin{tabular}{|c|c|c|c|c|}
\hline $B$ & $G$ & $\ci$ & $\ci/B^2$ & $E/B$ \\ \hline 1 & $O(3)$ &
1.0 & 1.000 & 1.232 \\ 2 & $D_{\infty h}$ & 5.8 & 1.452 & 1.208 \\ 3 &
$T_d$ & 13.6 & 1.509 & 1.184 \\ 4 & $O_h$ & 20.7 & 1.291 & 1.137 \\ 5
& $D_{2d}$ & 35.8 & 1.430 & 1.147 \\ 6 & $D_{4d}$ & 50.8 & 1.410 &
1.137 \\ 7 & $Y_h$ & 60.9 & 1.242 & 1.107 \\ 8 & $D_{6d}$ & 85.6 &
1.338 & 1.118 \\ 9 & $D_{4d}$ & 109.3 & 1.349 & 1.116 \\ 10 & $D_{4d}$
& 132.6 & 1.326 & 1.110 \\ 11 & $D_{3h}$ & 161.1 & 1.331 & 1.109\\ 12
& $T_d$ & 186.6 & 1.296 & 1.102\\ 13 & $O$ & 216.7 & 1.282 & 1.098\\
14 & $D_2$ & 258.5 & 1.319 & 1.103\\ 15 & $T$ & 296.3 & 1.317 &
1.103\\ 16 & $D_3$ & 332.9 & 1.300 & 1.098\\ 17 & $Y_h$ & 363.4 &
1.257 & 1.092\\ 18 & $D_2$ & 418.7 & 1.292 & 1.095\\ 19 & $D_3$ &
467.9 & 1.296 & 1.095 \\ 20 & $D_{6d}$ & 519.7 & 1.299 & 1.095 \\ 21 &
$T$  & 569.9  &  1.292   & 1.094 \\ 22 & $D_{5d}$ & 621.6 & 1.284  &
1.092\\ \hline
\end{tabular}
\caption{Results from the simulated annealing of rational maps of
degree $B$.  For $1\le B\le 22$ we list the symmetry of the rational
map, $G$, the minimal value of $\I$, its comparison with the bound
$\I/B^2\ge 1$,  and the energy per baryon $E/B$ obtained after computing
the profile function which minimizes the Skyrme energy functional.}
\label{tab-sa1}
\end{table}

\section{Skyrmion identification}\news
\label{sec-identify}

\begin{figure}
\vskip -3.0cm
\centerline{\epsfxsize=15cm\epsffile{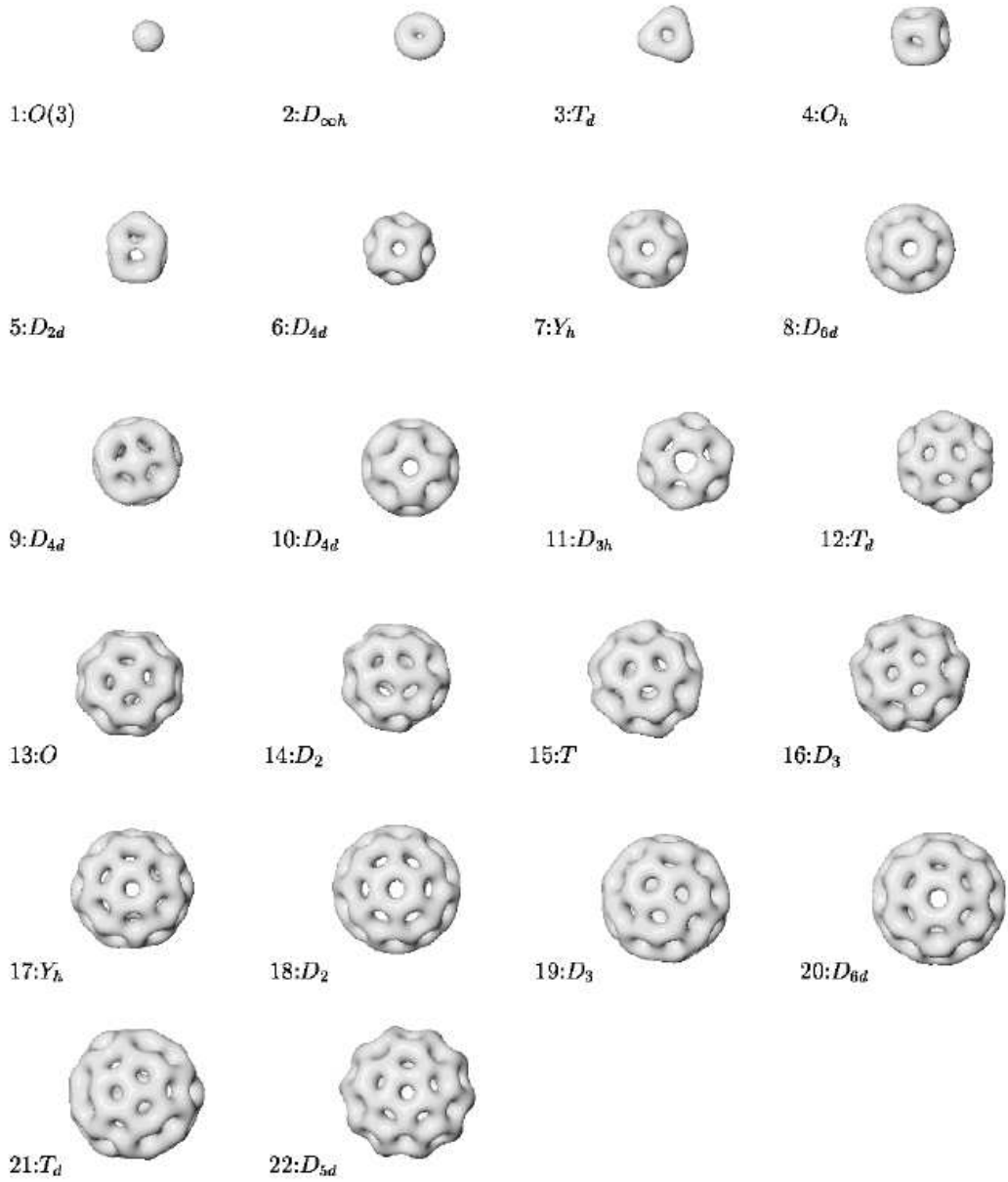}}
\vskip -0.0cm
\caption{The baryon density isosurfaces of the Skyrmions with $B=1-22$ which
 are minimum energy configurations (see table~\ref{tab-sa1}) within
 the rational map ansatz. Each corresponds to a value of ${\cal B}=0.035$ and are presented to scale.}    
\label{fig-minb}
\end{figure}

\begin{figure}
\vskip -3.0cm
\centerline{\epsfxsize=15cm\epsfysize=20cm\epsffile{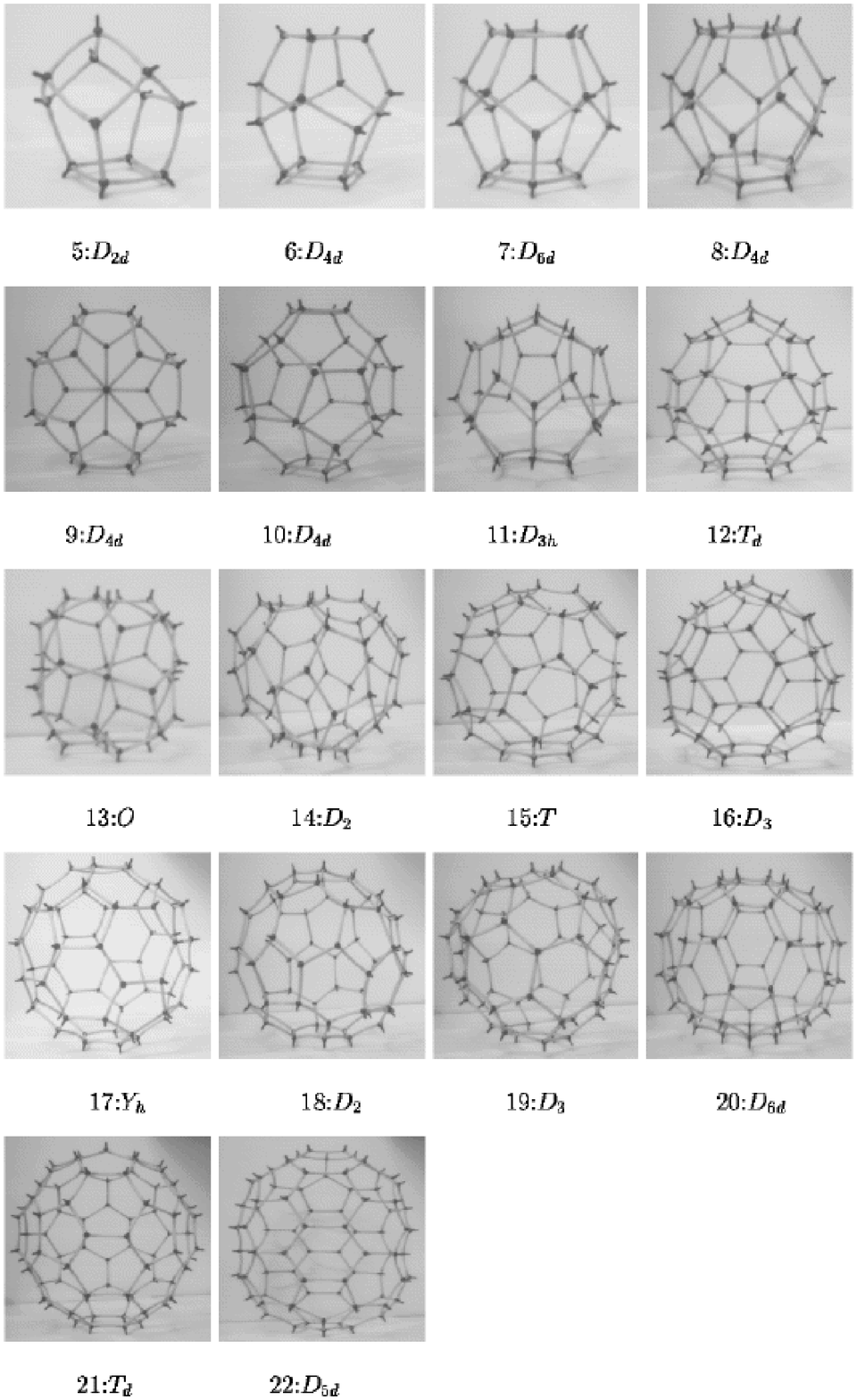}}
\caption{The associated polyhedra for the Skyrmions presented in fig.~\ref{fig-minb}. The models are not to scale.}    
\label{fig-minp}
\end{figure}

In this section we will present the results of an extensive set of
simulations performed with the intention of identifying the symmetry
and associated polyhedron of the minimum energy configurations for
$B\le 22$. We should note that making definitive statements as to an
identification for a particular charge has, historically, been fraught
with difficulties. In particular, the identifications of the $B=5$ and
$B=6$ solutions in ref.\cite{BTC} and the $B=9$ configuration in 
ref.~\cite{BS2} were incorrect to varying degrees. Suffice to
say when the two numerical methods agree for a wide range of initial
conditions and simulation parameters, there is strong grounds to
believe that we have created the correct configuration; identifying
the symmetry is then the only complication.  Conversely, when they
disagree, or we know more than one very low energy solution for a
particular charge, it is a matter of some debate which is the true
minimum energy solution, or whether there is some more complicated
symmetry allowing local, or even degenerate minima. We will
engage in this debate at various stages, but the reader should make up
their own mind as to the strength of our arguments.

We have argued strongly in the previous sections on numerical methods
that the simulated annealing of rational maps is the simplest and
probably cleanest approach to compute low energy Skyrmion
configurations. Clearly, its veracity for computing the true minima
depends on the viability of the ansatz for
describing Skyrmions, and that the energy functional based on $\ci$ is
a good approximation to the true energy, or at least the relative
energies of particular configurations. With these caveats in mind, we
present our first attempt at identification of the minima as the
rational maps which minimize $\ci$, before discussing other possibilities. 

The results of the simulated annealing algorithm applied to a general
rational map of degree $B(\le 22)$ and the symmetry identification
procedure discussed in section~\ref{sec-symmetric} are presented in
table~\ref{tab-sa1}. In each case, we tabulate the identified symmetry
group $G$, the minimum value of $\I$, the quantity $\I/B^2$ --- which
is strikingly uniform at around 1.2-1.3 --- and the value of $E/B$
for a profile function which minimizes the energy functional
(\ref{rmaenergy}) for the particular map.

We should first comment that for $B\le 8$ the rational maps which
minimize $\I$ are exactly those presented in ref.~\cite{HMS}  to
approximate the results of the full non-linear simulations~\cite{BS2};
thus the simulated annealing algorithm provides  a nice numerical
check that the same maps are reproduced by searching the full
parameter space of rational maps. Also for $B\ge 7$ all the symmetry
groups with the exception of $B=9$, $B=10$ and $B=13$ are compatible
with the fullerene hypothesis: that ${\cal S}_B$ has $4(B-2)$ trivalent
vertices and is constructed from $2B-14$ hexagons and 12
pentagons. The baryon density isosurfaces for each of the solutions
are displayed in fig.~\ref{fig-minb} along with a model of the
associated polyhedron in fig.~\ref{fig-minp},
 which confirm that for the most part 
they are indeed of the fullerene type. The symmetry groups of $B=9$,
$B=10$ and $B=13$ all contain the cyclic subgroup $C_4$, which is not
compatible with them being of the fullerene type, since the associated
polyhedron of such a
solution must contain either a four-valent bond, or a square. These
are the first Skyrme solutions found which do not comply with the
Geometric Energy Minimization (GEM) rules suggested in ref.~\cite{BS2}. We
shall discuss these solutions in more detail in the subsequent
section, but it is gratifying to note that all the other solutions
appear qualitatively  to comply with our expectations based on
the GEM rules.

\begin{figure}
\centerline{\epsfxsize=15cm\epsfysize=8cm\epsffile{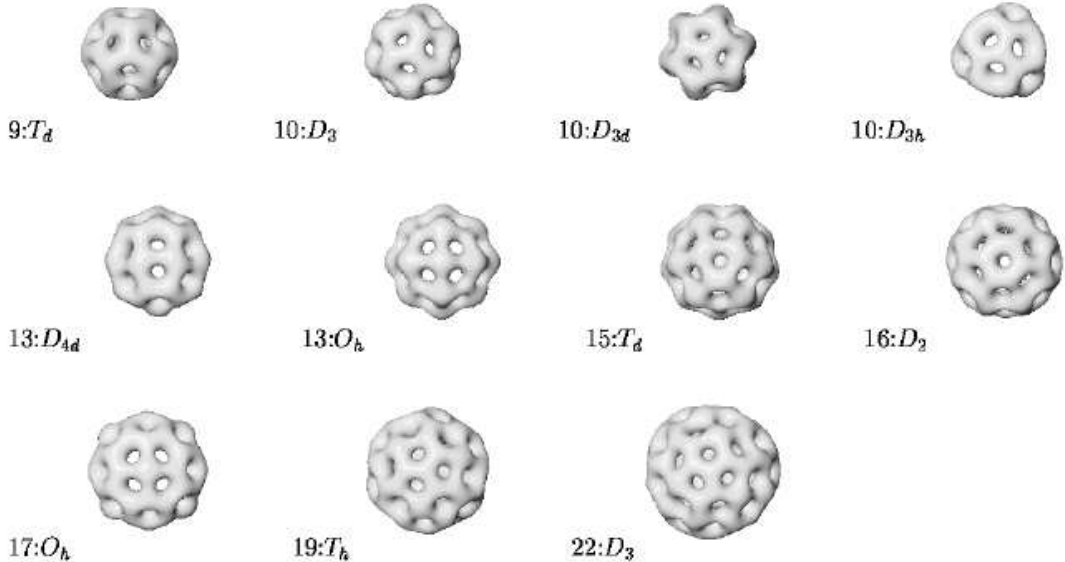}}
\vskip -0cm
\caption{The baryon density isosurfaces of the Skyrmions we found which are other 
stationary points of $\ci$ (see table~\ref{tab-sa2}) within the rational map ansatz. 
Each corresponds to a value of ${\cal B}=0.035$ and are presented to scale.}    
\label{fig-sadb}
\end{figure}

\begin{figure}
\vskip -0cm
\centerline{\epsfxsize=15cm\epsfysize=13cm\epsffile{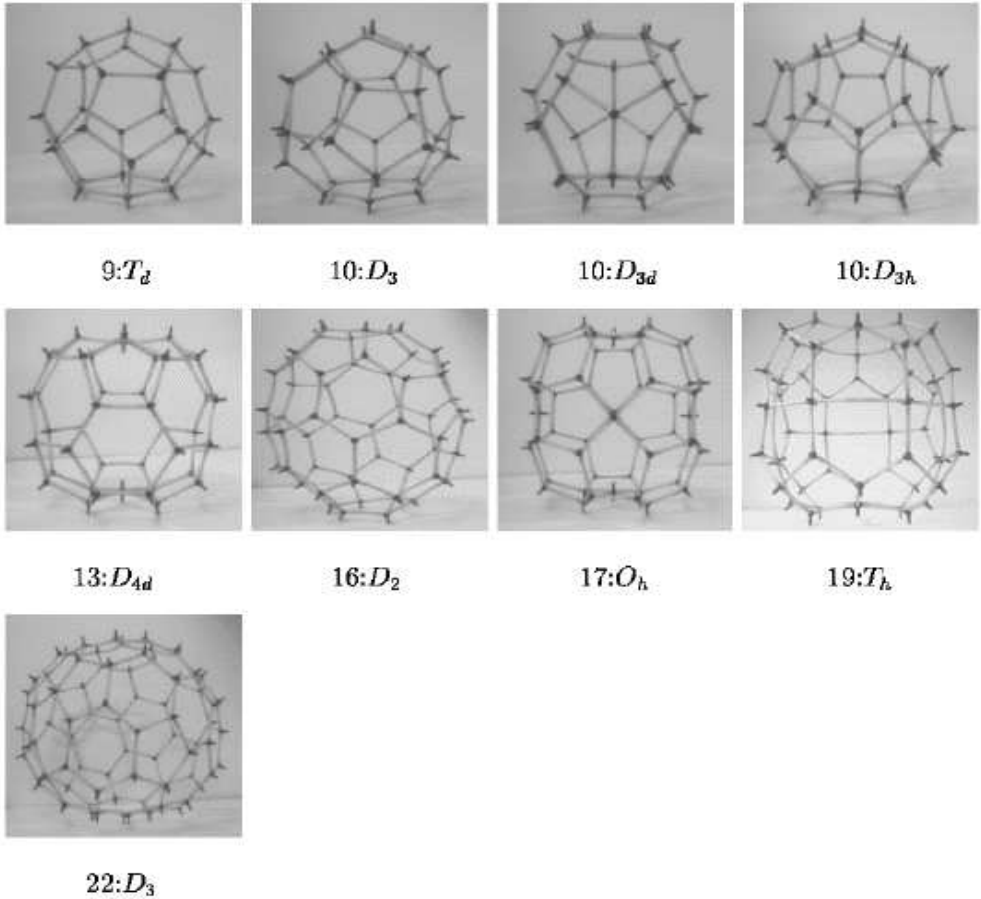}}
\vskip -0cm
\caption{The associated polyhedra for the Skyrmions presented in fig.~\ref{fig-sadb}. The models once again are not to scale. Note that we have been unable to make models for the $B=13$ solution with $O_h$ symmetry and the $B=15$ solution with $T_d$ symmetry since they contain a large number of four-valent bonds. }
\label{fig-sadp}
\end{figure}

In table~\ref{tab-sa2}, fig.~\ref{fig-sadb} and fig.~\ref{fig-sadp}, 
we present the results of our extensive search for the
minimum energy maps with particular symmetries, usually dihedral groups
or chosen from the extensive tables of fullerenes presented in
ref.~\cite{atlas}, which lend further weight to our
conclusions that those presented in table 1 are in fact the minima
relative to the energy functional $\I$. They do, however, turn up the
possibility that in certain cases the minima of $\I$ may not
necessarily be the minimum of the Skyrme energy, since some of them
have values of $\I$ very close to the values presented in
table 1. When the values are so close it is difficult to make a guess as
to how the relaxation to the true solution might effect their relative
positions; an issue to which we shall return in subsequent sections. For
the moment we will denote them by $\star$, and conclude at least that
they are not a global minima of $\ci$, but are believed to represent other
critical points.

For $9\le B \le 22$  we shall now describe in detail the rational maps we
have obtained, the structure of the associated Skyrmions
and make a comparison with the results from full field
simulations. Our study of the rational maps has already turned up a
few oddities and we shall attempt to interpret these at the relevant
charge. Charges where the fullerene hypothesis appears to break down
are $B=9$ and $B=13$, while the rational map approach to representing
the minimum energy Skyrmion appears to need careful consideration for $B=10$,
$B=14$, $B=16$ and $B=22.$

\begin{table}
\centering
\begin{center}
\begin{tabular}{|c|c|c|c|c|}
\hline $B$ & $G$ & $\ci$ & $\ci/B^2$ & $E/B$ \\ \hline
 9* & $T_d$ & 112.8 & 1.393 & 1.123\\ 10* & $D_3$ &
132.8 & 1.328 & 1.110\\ 10* & $D_{3d}$ & 133.5 & 1.335 & 1.111 \\ 10*
& $D_{3h}$ & 143.2 & 1.432 & 1.126\\ 13* & $D_{4d}$ & 216.8 & 1.283 &
1.098\\ 13* & $O_h$ & 265.1 & 1.568 & 1.140\\ 
15* & $T_d$ & 313.7 & 1.394 & 1.113\\
16* & $D_{2}$ & 333.4 & 1.302 & 1.098\\
17* & $O_h$ & 367.2  & 1.271 & 1.093\\
19* & $T_h$  & 469.8 & 1.301 &  1.096\\ 
22* & $D_3$  & 623.4 & 1.288 &  1.092 \\
\hline
\end{tabular}
\end{center}
\caption{Same as for table~\ref{tab-sa2}, but for the other critical
points of $\ci$. Notice that the $\ci$ values for the $B=10$
configurations with $D_3$ and $D_{3d}$ symmetry, the $B=13$ with
$D_{4d}$, the $B=16$ with $D_2$ and the $B=22$ with $D_3$  are
extremely close to the corresponding values in table~\ref{tab-sa1},
suggesting the possibility of local minima.}
\label{tab-sa2}
\end{table}

\subsection{$B=9$}

In ref.~\cite{BS2} it was suggested that the $B=9$ minimum energy
configuration had $T_d$ symmetry, a symmetric configuration of ${\rm
C}_{28}$ (it corresponds to configuration 28:2 in
ref.~\cite{atlas}). The polyhedron to which it corresponds comprizes
of 12 pentagons, fuzed into 4 triplets, placed at the vertices of a
tetrahedron, with  4 hexagons placed at the vertices of a dual
tetrahedron. This, plus the solutions for $B=7$ and $B=8$, was one of
the main motivations of the fullerene hypothesis. Unfortunately, it
appears that the original identification of the symmetry was incorrect
and further relaxation of this configuration using the full non-linear
field equations lead to a somewhat different solution.
 
The minimizing map in this case has $D_{4d}$ symmetry and $\ci=109.3$,
with the functional form of the map being given by \be
R=\frac{z(a+ibz^4+z^8)}{1+ibz^4+az^8}\,,
\label{9a}
\ee where $a=-3.38, b=-11.19.$ It should be noted that this is
slightly lower than the value $\ci=112.8$ of the tetrahedral
map~\cite{HMS}  \be
R=\frac{5i\sqrt{3}z^6-9z^4+3i\sqrt{3}z^2+1+az^2(z^6-i
\sqrt{3}z^4-z^2+i\sqrt{3})}{z^3(-z^6-3i\sqrt{3}z^4+9z^2-5i
\sqrt{3})+az(-i\sqrt{3}z^6+z^4+i\sqrt{3}z^2-1)}\,,
\label{9b}
\ee where $a=-1.98.$

Amazingly, the solution which was created by the relaxation of
initially well-separated  $B=8$ and $B=1$ configurations  (henceforth
such initial conditions will denoted $8+1$)  using the full non-linear
field equations, and confirmed using a number of different initial
conditions (for example, $7+2$ and $6+3$), is precisely that which
corresponds to the rational map (\ref{9a}). The associated polyhedron
is not a fullerene, since the symmetry group $D_{4d}$ is incompatible
with pentagons and hexagons forming a trivalent polyhedron.  In fact,
it has two four-valent links which occur between four pentagons
forming the top and bottom pseudo-faces\footnote{We shall use the term
pseudo-face to refer, rather loosely, to a set of connected polygons,
which act from the point of view of symmetry of the associated polygon
as a single face.}
 of a rather flat polyhedron, linked by a
belt of eight alternately up and down pointing pentagons; the top and
bottom being rotated relative to each other by $45^{\circ}$ to give
the $D_{4d}$ symmetry. In section~\ref{sec-enhance} we will discuss
how this solution can be formed by the symmetry enhancement of the
other known fullerene corresponding to ${\rm C}_{28}$ which has $D_2$
symmetry (this is labelled 28:1 in ref.~\cite{atlas}).
On the basis of this we conclude that ${\cal S}_9$ has
$D_{4d}$ symmetry and not $T_d$ as previously suggested, but that the
known $T_d$ symmetric solution is a saddle point.

\subsection{$B=10$}

A glance at table~\ref{tab-sa1} and table~\ref{tab-sa2} shows that
there  are (at least) four maps whose $\I$ values are very close
together. Minimizing over all maps yields the value $\ci=132.6$ and
this can be obtained from a $D_{4d}$ symmetric map of the form \be
R=\frac{z^2(a+ibz^4+z^8)}{1+ibz^4+az^8}\,,
\label{10_d4d}
\ee where $a=-8.67$ and $b=14.75.$ This is not a fullerene Skyrmion,
the four-fold symmetry this time manifesting itself in the existence
of a square on the top and bottom of the associated polyhedron.

There is, however, a fullerene Skyrmion with $\I=132.8$, very close to
that of the minimum, which corresponds to a $D_3$ symmetric map of the
form \be R=\frac{z(1+az^3+bz^6+cz^9)}{c+bz^3+az^6+z^9}\,,
\label{10}
\ee where $a,b,c$ are complex parameters.  The minimum is obtained for
the values  $a=4.40-1.72i, b=-2.38+3.10i, c=-0.12+0.19i.$ The symmetry
can be increased to $D_{3d}$ by choosing $b$ to be real and $a$ and
$c$ to both be purely imaginary, and within this class the minimum is
very slightly higher at $\ci=133.5,$ which is attained when $a=20.40i,
b=-30.22, c=-4.69i.$ Finally, if $a,b,c$ are all real then the
symmetry is $D_{3h}$ and the minimum in this class has $\ci=143.2$
when $a=-5.14, b=-2.20, c=-0.36.$

The baryon density of the $D_{4d}$ symmetric map is presented in
fig.~\ref{fig-minb}
and for the other three maps, $D_{3},D_{3d},D_{3h}$, in
fig.~\ref{fig-sadb}.
The polyhedron associated with the $D_{4d}$ solution  can be constructed
by taking two squares each surrounded by 4 hexagons,
connected via a band of 8 pentagons alternately pointing up and down;
the two squares being rotated relative to each other by
$45^{\circ}$. Each of the links is trivalent, but instead of
comprizing of 12 pentagons and 6 hexagons as would have been suggested
by the fullerene hypothesis, it contains 8 pentagons, 8 hexagons and 2
squares, although the number of vertices  $32\equiv 4(B-2)$ and
faces $18\equiv 2(B-1)$ are still compatible with the GEM rules  The
other three maps give Skyrmions of fullerene type, with the baryon
density isosurface comprising the requisite number of  pentagons and
hexagons arranged in a trivalent polyhedron. The associated polyhedron
for the $D_{3h}$ solution (which corresponds to 32:5 in ref.~\cite{atlas})
comprizes of two copies of a hexagonal triple linked by a belt of 12
pentagons which can be thought of as being made of 3 sets of four
fuzed pentagons in a $C_3$ arrangement.  The $D_3$ and $D_{3d}$
solutions are very similar: to make each of the associated polyhedra
(which correspond to configurations 
32:6 and 32:4 in ref.~\cite{atlas} respectively) 
first start with two pentagon
triples. There are six places on each triple to which one can add
another polygon, which fall into two types --- one can connect to a
single pentagon edge, or between two pentagons connecting to an edge
of each. To make the two different configurations, one must add three
hexagons and three pentagons alternately around each of the triples;
the difference being which polygon (pentagon or hexagon) connects to
the two different sites. In particular, the $D_{3d}$ configuration has
hexagons connected to the single pentagon, and pentagons between the
two pentagons;  vice versa for the  $D_3$ configuration. Once one has
added these 6 polygons, the two identical copies are then connected,
the two being rotated relative to each other by $60^{\circ}$ in the
$D_{3d}$ configuration, and at no particular fixed angle in the case
of $D_3$.  As fullerenes these three structures are tabulated  in
ref.~\cite{atlas}, along with three other possibilities which have
less symmetry ($2\times C_2$ and $D_2$). Unfortunately, these lower
symmetry solutions are impossible to find using the simulated annealing
algorithm since their symmetry groups are subgroups of $D_{4d}$ and
the minimum energy rational map with this symmetry is already known.

In order to try and understand which of the four configurations is the
true minimum we have relaxed all the different initial configurations
made from two individual Skyrmion clusters whose baryon numbers sum to
10, that is, $9+1$, $8+2$, $7+3$, $6+4$ and $5+5$. None of these form
the $D_{4d}$ configuration, nor that with $D_{3h}$, suggesting --- but
not proving --- that neither of them are the minimum energy solution,
and are hence likely to be saddle points.  However, it appears that it
is possible to produce both the  $D_{3d}$ and  $D_3$ configurations
from collisions. In particular, the $7+3$ relaxation appears to give
the more symmetric $D_{3d}$ configuration, while all the others give
one which only has $D_3$ symmetry.  We have already found solutions
which we believe to be saddle points, for example, the $B=9$
configuration with $T_{d}$ symmetry, but here we have evidence for a
new phenomena --- local minima. Given that there is no symmetry which
can be invoked to explain why there might be degenerate minima, it
seems likely that the energies of the two configurations must differ
by a minute amount. As we shall discuss in section~\ref{sec-energies},
given the uncertainties we are unable to ascertain which is the global
minimum. 

The reason that the rational map ansatz is so successful in describing
Skyrmions is that they appear to prefer to be as spherical\footnote{A
sensible  quantification of how spherical a solution is might be to
consider the eigenvalues of the moments of the baryon density. A
distribution which is isotropic, and hence almost spherical, would
have all the eigenvalues approximately equal, whereas a more elongated
solution would have one which is substantially different to the other
two.} as possible. Examination of the models of the associated
polyhedra sheds some light on the preference of the rational map ansatz
for the $D_{4d}$ and $D_3$ configurations, which are very spherical,
as opposed to the more elongated $D_{3d}$ configuration. It might be
that this oddity is not reproduced in the full non-linear energy
functional, and the configuration with $D_{4d}$ symmetry is impossible
to reproduce.

The phenomenon of many different configurations at a given baryon
number, often with energies very close to the minimal value, is a
feature of the fullerene hypothesis which one might have been able  to
predict since there are many possible polyhedra which contain 12
pentagons and $2B-14$ hexagons for $B\ge 9$. The possibility of
four-valent vertices and also trivalent configurations containing
squares only make things worse. This was the main motivation for the
choice of a simulated annealing algorithm as our minimization scheme
for rational maps; a choice which appears to have been vindicated.  It
is also one reason why we have used two very different numerical
techniques, the rational map approach and full field simulations, to
try and confirm the results we obtain, thereby increasing the
confidence that the solutions we construct are the global
minima. Unfortunately, in this case we were unable to make a
definitive identification of the minimum energy Skyrmion, ${\cal
S}_{10}$,  but have presented evidence that it is one of two
configurations which are almost indistinguishable.

\subsection{$B=11$}

The minimum value at this charge is $\ci=161.1$ and this is obtained
from the $D_{3h}$ symmetric map \be
R=\frac{z^2(1+az^3+bz^6+cz^9)}{c+bz^3+az^6+z^9}\,,
\label{11}
\ee where $a,b,c$ are real parameters, taking the values
$a=-2.47,b=-0.84,c=-0.13$ at the minimum. The associated polyhedron
(which corresponds to 36:13 in ref.~\cite{atlas})
can be constructed by considering a hexagon to which 3 pentagons and 3
hexagons are connected alternately with a $C_3$ symmetry. Each of the
pentagons is part of a set of four fuzed together, each of which 
is placed in the $C_3$
arrangement. Each of these is then connected to another hexagon, which
is directly below the first one and can be thought of as being
equivalent to the original one from the point of view of symmetry. The
spaces in between are filled up with hexagons, the whole structure
comprizing of 12 pentagons and 8 hexagons. 

The exact same configuration was produced by the collision and then
relaxation of two Lorentz-boosted $B=3$ Skyrmions and a stationary
$B=5$ solution in a linear arrangement ($3+5+3$).  Given that the
fullerene hypothesis is clearly not the whole story for $B=9$ and
$B=10$, it is reassuring that things get back on track at $B=11$ with
what appears to be the unique global minimum, ${\cal S}_{11}$, being a
fullerene type solution describable by the rational map ansatz.

\subsection{$B=12$}

Considering all degree 12 maps the minimum is found to be $\ci=186.6$
and this can be reproduced from a $T_d$ symmetric map constructed as
follows.  Decomposing $\underline{13}$ as a representation of $T$
gives \be\underline{13}\vert_T=2A+A_1+A_2+3F.\ee Now let $p_{\pm}$ be the
Klein polynomials \cite{Kl} \be p_\pm=z^4\pm2\sqrt{3}iz^2+1\,, \ee
associated with the vertices and faces of a tetrahedron.  On applying
the $C_3$ generator contained in the tetrahedral group to these
polynomials they acquire the multiplying factors  $p_\pm\mapsto
e^{\pm2\pi i/3}p_\pm.$ Thus, the degree 12 polynomials $p_\pm^3$ are
strictly invariant, forming a basis for the representation $2A$ in the
above decomposition and the polynomials $p_+^2p_-$ and  $p_+^2p_-$ are
bases for the representations $A_1$ and $A_2$ respectively.
Explicitly, the rational map  \be R=\frac{ap_+^3+bp_-^3}{p_+^2p_-}\,,
\label{12}
\ee is $T_d$ symmetric for all real $a$ and $b$, with the minimal
value $\ci=186.6$ obtained for $a=-0.53, b=0.78.$ We should note that
there are other maps with $T_d$ symmetry (the denominator in
(\ref{12}) can be replaced by $p_+^3$, for example), but it appears
that  all these have a larger value for $\ci.$

As for $B=11$, this fullerene-like configuration was reproduced in
non-linear field theory simulations, this time from initially
well-separated $B=7$ and $B=5$ solutions ($7+5$), allowing us to
conclude that it is the unique ${\cal S}_{12}$. The associated
polyhedron (which corresponds to configuration 40:40 in ref.~\cite{atlas})  
is in some ways similar to the $T_d$ solution at $B=9$:
there being four pentagon triplets positioned on the vertices of a
tetrahedron. Each of these triplets is completely surrounded by
hexagons  forming a  polyhedron well-known in fullerene
chemistry~\cite{atlas}, where it is one of 40 configurations with 12
pentagons and 10 hexagons which are candidates for a ${\rm C}_{40}$
cage.

\subsection{$B=13$}

The minimal map of degree 13, deduced from simulated annealing of
general maps, has cubic symmetry, another with four-fold symmetry
which is incompatible with the fullerene hypothesis. It is
interesting to note that the fullerene hypothesis would have predicted a
trivalent polyhedron made from 12 pentagons and 12 hexagons. We have not
discussed representations of the cubic group\footnote{The cubic
symmetry group $O$ is that of the octahedron/cube, without all the
reflection symmetries contained in the full symmetry group $O_h$.}, $O$, so we shall describe
the group theory of this example by embedding it into the tetrahedral
group, whose representations were reviewed earlier. 

One finds that  \be  \underline{14}\vert_T=3E'+2E_1'+2E_2'\,,
\label{decomp14}
\ee  so there is a two parameter family of $T$ maps associated with
the first component in (\ref{decomp14}). Setting one of these
parameters to zero extends the symmetry to $O$ and results in the one
parameter family of maps  \be
R=\frac{z(a+(6a-39)z^4-(7a+26)z^8+z^{12})}{1-(7a+26)z^4+(6a-39)z^8+az^{12}}\,,
\label{13_O}
\ee whose minimum occurs at $a=0.40+5.18i$ when $\I=216.7.$  The
associated polyhedron is in many ways similar to a cube comprizing of
six pseudo-faces, each of which are made of four
pentagons with a four-valent bond, very similar to those in the  $B=9$
configuration.  Clearly, in order for the
them to fit together, with all the other bonds being trivalent,
each of these pseudo-faces must be rotated slightly 
 relative to the one diametrically
opposite, which  removes the possibility of the reflection symmetries
of the cube and, hence, the symmetry group $O_h$. The
polyhedron comprizes of a total of 24 pentagons, as opposed to the 12
pentagons and 12 hexagons that would have been expected had the
fullerene hypothesis been correct for this charge. As we shall discuss in
section~\ref{sec-enhance}, this solution, which is reproduced in
relaxation of a range of initially well separated clusters ($3+7+3$
and $12+1$, for example), is rather special; it being obtainable via a
multiple symmetry enhancement of a $D_2$ fullerene polyhedron
(probably from either configurations 44:75 or 44:89 in ref.~\cite{atlas}).

If $a$ is real then the symmetry can be extended to $O_h$, but the
minimum in this class is quite a bit higher at $\I=265.1$ for $a=7.2.$
This  Skyrmion, which is probably a saddle point, has recently
been computed in ref.~\cite{MP} from the relaxation of a single
Skyrmion surrounded by 12 others in an initially face centred cubic
array. The polyhedron associated with this configuration is more akin
to the octahedron than the cube, comprizing of eight triangular pseudo-faces.  It contains mainly four-valent bonds; the only trivalent ones
being placed in the centre of the pseudo-faces.

There is a degree 13 map with $D_{4d}$ symmetry whose $\I$ value is
extremely close to the minimal one, in fact $\I=216.8.$ This map is
\be
R=\frac{z(ia+bz^4+icz^8+z^{12})}{1+icz^4+bz^8+iaz^{12}}
\label{13}
\ee where $a,b,c$ are real and take the values $a=-5.15, b=-50.46,
c=46.31$ at the minimum. This Skyrmion looks similar to the $O$
symmetric minimum, 
but it only has two four-valent vertices as opposed to the six in the
cubic configuration, and can be thought of as being an extension of  the
$D_{4d}$ configuration with $B=9$. The two four valent bonds are part
of two pseudo-faces forming the top and bottom, which are linked by 
eight copies of a single hexagon connected to a pentagon, alternately arranged
pointing up and down, so that four hexagons and four pentagons connect
to both the top and bottom pseudo-faces.
 This configuration contains 8 hexagons and
16 pentagons, breaks the GEM rules since the number of vertices is 42
rather than the predicted number 44 
(the number of faces is still $24\equiv 2(B-1)$) and can be created by a
single symmetry enhancement of the same $D_2$ fullerene as the $O$
symmetric configuration (see section~\ref{sec-enhance} for details).
 Given the similarity of this configuration
to the minimum, it is  no surprize that the values of $\I$ are very
close. We conclude, therefore, that  ${\cal S}_{13}$ has $O$
symmetry, can be approximated by the rational map (\ref{13_O}),
 and that the GEM rules and the fullerene hypothesis breakdown at this
charge as they did for $B=9$.

\subsection{$B=14$}

The minimizing map of degree 14 has only a relatively small symmetry,
that of $D_2.$ The map can be written in the form 
\be
R=\bigg(\sum_{j=0}^7a_jz^{2j}\bigg)/\bigg(\sum_{j=0}^7a_{7-j}z^{2j}\bigg)\,,
\label{14}
\ee
where $a_7=1$ and $a_0,..,a_6$ are complex parameters.  The minimum is
$\ci=258.5$ which occurs when the parameters are those given in
table~\ref{tab-coeff14}. This configuration has much less symmetry
than any of the others previously described and the associated
polyhedron is difficult to visualize in detail. It can be constructed by
arranging the 12 pentagons in 4 sets of 3. The 4 sets should be split
up into two pairs, each of which is connected by three hexagons, one
in the gap between two pentagons on each side, and the other two
either side of the first.  The two pairs should then be connected by a
band of a further 8 hexagons, making 14 in total. The configuration is
of the fullerene type (it corresponds to configuration 48:144 in
ref.~\cite{atlas}) and is one of 192 possibilities containing 12
pentagons and 14 hexagons.

\begin{table}
\centering
\begin{center}
\begin{tabular}{|c|c|c|c|c|c|c|c|}
\hline & $a_0$ & $a_1$ & $a_2$ & $a_3$ & $a_4$ & $a_5$ & $a_6$ \\
\hline ${\rm Re}(a)$ & 0.8 & -5.0 & -3.0 & -53.4 & -15.2 &  -13.1 &
0.9 \\ \hline ${\rm Im}(a)$ & 0.3 & -13.5 & 3.7 & -59.4 & 66.2 &  34.1
& -11.6 \\ \hline
\end{tabular}
\end{center}
\caption{The coefficients of the  minimal $D_2$ map with $B=14$.}
\label{tab-coeff14}
\end{table}

\begin{figure}
\vskip -0cm
\centerline{\epsfxsize=15cm\epsfysize=6cm\epsffile{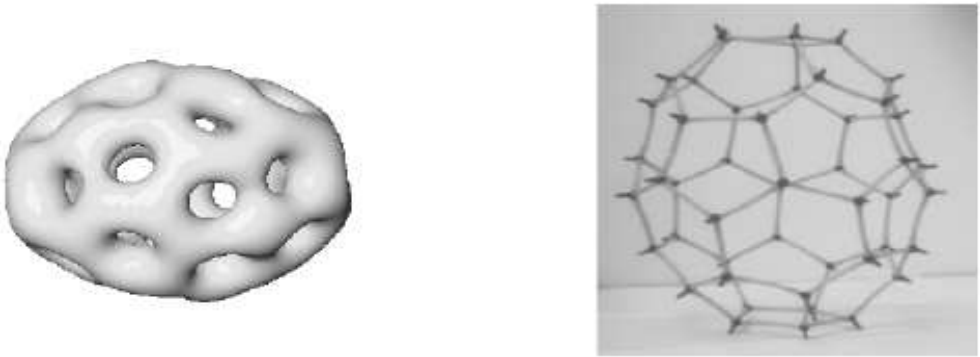}}
\vskip -0cm
\caption{The baryon density isosurface and associated polyhedron of
the $B=14$ solution with $C_2$ which is created during the collision
of well-separated Skyrmion clusters. We believe this elongated
solution to be the minimum energy Skyrmion at this charge.}
\label{fig-charge14}
\end{figure}

Attempts to reproduce this solution by the relaxation of
well-separated clusters have proved unsuccessful. We have tried
initial conditions which comprize of $7+7$, $12+2$ and $13+1$ and in
each case the same configuration, shown in fig.~\ref{fig-charge14} was the
end-product. This configuration has even less symmetry than the
minimum energy rational map; it having just $C_2$ symmetry. The
associated polyhedron is almost impossible to describe due to the lack
of symmetry suffice to say that it contains 12 pentagons and 14
hexagons, and corresponds to configuration 48:83\footnote{This was
identified by computing the pentagon index for the associated
polyhedron and checking it against the table in ref.~\cite{atlas}.} in
ref.~\cite{atlas}.

Note that this Skyrmion is very elongated and so it is not surprising
that the rational map approximation does not describe this
configuration very well, since it assumes that the baryon density has
the same angular distribution on  concentric spherical
shells. Presumably there is a rational map which describes a
distorted, more spherical, version of this Skyrmion, but its $\I$
value will be larger than the minimal one. Unfortunately we are unable
to find this rational map using simulated annealing 
since its symmetry group is contained within that of the minimizing map. 
We believe that ${\cal S}_{14}$ is the elongated
configuration shown in fig.~\ref{fig-charge14}, based on the fact that it
was created easily in the collision of well-separated clusters. It is
of the fullerene type,  but at this stage we do not have a good description of
it in terms of a rational map. In subsequent sections, when we use
the rational map ansatz as the starting point for a quantitative
investigation of Skyrmion properties we will be forced to use the
$D_2$ configuration instead of what we believe to be the
minimum. However, this should not lead to substantial errors.

\subsection{$B=15$}

Considering all degree 15 maps the minimum is found to be $\ci=296.3$
which has tetrahedral symmetry\footnote{As for the cubic group $O$,
the group $T$ is that of the tetrahedron, but without the reflection
symmetries of $T_d$.}, $T$.  To construct the map the relevant
decomposition is \be
\underline{16}\vert_T=2E'+3E_1'+3E_2'.\label{decomp16}\ee At first
sight it may appear that there is a one (complex) parameter family of
tetrahedral maps corresponding to the $2E'$ component. However, this
is not the case since this family of maps is degenerate, having common
factors. From  \be
\underline{8}\vert_T=2E'+E_1'+E_2'\,,\label{decomp8}\ee it follows
that there is a one parameter family of degree seven tetrahedral maps
(this family is constructed explicitly in \cite{HMS}).  Furthermore,
\be \underline{9}\vert_T=A+A_1+A_2+2F\,,\ee so there is a strictly
invariant degree eight tetrahedral polynomial, which is given by
$p_+p_-=1+14z^4+z^8$ and is the vertex polynomial of a cube.  A basis
for the $2E'$ component of (\ref{decomp16}) is obtained by multiplying
each basis polynomial for the $2E'$ component in (\ref{decomp8}) by
$p_+p_-$, and hence the corresponding map is degenerate, being only
degree seven rather than 15.

The $3E_1'$ component in (\ref{decomp16}) does correspond to a genuine
two (complex) parameter family of degree 15 tetrahedral maps. Using
the methods described in ref.~\cite{HMS} we find that this family of
maps is given by $R=p/q$ where
\begin{eqnarray}
p= &&i\sqrt {3}\left (1+a-b\right ){z}^{15}+\left (77-99\,a-5\,b
\right ){z}^{13} \nonumber\\ &+& i\sqrt {3}\left
(637+21\,a+35\,b\right ){z}^ {11} +\left (1001+561\,a-65\,b\right
){z}^{9} \nonumber\\ &+& i\sqrt {3}\left  (-429+99\,a+45\,b\right
){z}^{7}+\left (-1001-297\,a-127\,b\right ){z} ^{5} \nonumber\\
&-&i\sqrt {3}\left (273+185\,a+15\,b\right ){z}^{3}+\left (
115+27\,a+5\,b\right )z\,,
\end{eqnarray}
and $q(z)=z^{15}p(1/z)$. The value $\ci=296.3$ is obtained when
$a=0.16+2.06i$, $b=-4.47-8.57i.$ If $a$ and $b$ are both real then the
symmetry extends to $T_d$, but the minimum in this class is higher at
$\ci=313.7,$  when $a=4.64, b=-20.45.$

The polyhedron associated with the $T$ symmetric solution is of the
fullerene type. It contains 12 pentagons and 16 hexagons which can be
thought of as being arranged in 8 pseudo faces: 4 of these comprize of
hexagon triples, whereas the other 4 can be made from a hexagon
connected to 3 pentagons in a $C_3$ arrangement. The 4 hexagon triples
can be thought of as being placed on the vertices of a tetrahedron,
and the other 4 pseudo-faces, which are connected to the others, can
be thought of as being on the vertices of another tetrahedron which
is not dual to the first, removing the possibility of reflection
symmetries. 

There is no fullerene polyhedron with $T_d$
symmetry~\cite{atlas}, and hence this configuration must contain
four-valent bonds. In fact it contains more four valent bonds than
trivalent ones, in an essentially similar way to the $O_h$
configuration for $B=13$.  The $T$ symmetric solution was reproduced
in the relaxation of clusters containing $12+3$ and $13+2$, initially
well separated, and hence we conclude that ${\cal S}_{15}$ has $T$
symmetry, is of the fullerene type and can be reproduced by a rational
map.

\subsection{$B=16$}

$B=16$ is another interesting situation where it appears that the
minimum energy rational map may not in fact be the minimum energy
Skyrmion.  The minimizing map of degree 16 has $D_3$ symmetry which
takes the form \be
R=\bigg(\sum_{j=0}^5a_jz^{3j+1}\bigg)/\bigg(\sum_{j=0}^5a_{5-j}z^{3j}\bigg)\,,
\label{16a}
\ee where $a_5=1$ and $a_0,..,a_4$ are complex parameters.  The
minimum is $\ci=332.9$ which is attained when the parameters take the
values presented in  table~\ref{tab-coeff16a}. The associated
polyhedron can be constructed by first taking two sets of 3 hexagons,
each of which is almost flat. Now connect a total of 6 pentagons and
3 hexagons, in 3-fold cyclic order around the flat structure, such
that each of the gaps between the original hexagons is filled by a
pentagon flanked on one side by another pentagon and
on the other side by a
hexagon. A further 6 pentagons, split into 3 pairs are used to fuze
the two halves; each of the pairs connecting sets of 2 pentagons.
Within the structure, there are a number of symmetrically placed
groupings of polygons which are exactly those which 
were found to lead to symmetry enhancement as observed
for $B=9$ and $B=13$. However, in this case a close examination of the
baryon density isosurface shows that all the bonds remain trivalent
and the associated 
polyhedron is of the fullerene type, containing 12 pentagons
and 18 hexagons.

\begin{table}
\centering
\begin{center}
\begin{tabular}{|c|c|c|c|c|c|}
\hline & $a_0$ & $a_1$ & $a_2$ & $a_3$ & $a_4$  \\ \hline ${\rm
Re}(a)$& 5.4 &-14.6 &  35.9 & -125.2 &  -5.2 \\ \hline ${\rm Im}(a)$&
-0.4 &-69.3 & 165.9 & 77.4 & 34.2 \\ \hline
\end{tabular}
\end{center}
\caption{The coefficients of the  minimal $D_3$ map with $B=16$.}
\label{tab-coeff16a}
\end{table}

There exists a family of $D_2$ symmetric rational maps with $B=16$ of
the form  \be
R=\bigg(\sum_{j=0}^8a_jz^{2j}\bigg)/\bigg(\sum_{j=0}^8a_{8-j}z^{2j}\bigg)\,,
\label{16b}
\ee where $a_8=1$ and $a_0,..,a_7$ are complex parameters. The minimum
in this class takes place when $\ci=333.4$, very close to that with
$D_3$, and the parameters take the values presented in
table~\ref{tab-coeff16b}. Since the solution has very little
symmetry it is difficult to describe the associated polyhedron as for
that with $B=14$. It is
of the fullerene type, comprizing of 12 pentagons and 18 hexagons, and
can be thought of being formed from two identical half shells,
connected together. To construct each of the two shells, start with a
hexagon and attach to it, in cyclic order, two pentagons, a hexagon,
another two pentagons and another hexagon. Then connect a pentagon,
pointing downwards to the edge of the two hexagons, and add three
hexagons to each side connecting the two pentagons.

\begin{table}
\centering
\begin{center}
\begin{tabular}{|c|c|c|c|c|c|c|c|c|}
\hline & $a_0$ & $a_1$ & $a_2$ & $a_3$ & $a_4$ & $a_5$ & $a_6$ & $a_7$
\\ \hline ${\rm Re}(a)$& 0.0&-4.2&6.9&39.8&-76.4&-201.0& -5.9&-9.7\\
\hline ${\rm Im}(a)$& 0.5&19.9&4.2&-105.0&64.8&-41.0&27.8&-2.7\\
\hline 
\end{tabular}
\end{center}
\caption{The coefficients of the  minimal $D_2$ map with $B=16$.}
\label{tab-coeff16b}
\end{table}

Given that there are at least two rational maps with very similar
values of $\ci$, it is interesting to see if we can create them both
via the relaxation of initial well-separated clusters. To this end we
have tried a number of different initial conditions ($7+9$, $12+4$ and
$13+3$). In contrast to the $B=10$ case where we were able to create
both the $D_3$ and $D_{3d}$ configurations, in each case the
end-product of the relaxation process had $D_2$ symmetry. This
strongly suggests, but does not prove, that this is the global minimum
energy solution and that the $D_3$ configuration may be a
saddle point solution. This is interesting since it suggests
that the relative ordering of the $D_3$ and $D_2$ solutions is probably
different when considering the full non-linear energy functional and
that for the rational map ansatz.

\subsection{$B=17$}

The case of $B=17$ is interesting since it was conjectured in
ref.~\cite{BS2} on the basis of the fullerene hypothesis that there
might exist a Skyrmion configuration of this charge
with the same structure as that
of Buckminsterfullerene, ${\rm C}_{60}$, which comprizes of 20
hexagons and 12 pentagons in an icosahedral configuration. This is
well known as the standard design for a football since it is almost
spherical, and also in civil engineering where it was championed by
Buckminster Fuller as a candidate for a geodesic dome. It is the
isolated pentagon structure (each pentagon is isolated by connecting
it to 5 hexagons) with the lowest number of vertices, and appears to
be the most stable carbon cage; it being the subject of much interest
in chemistry in the recent past. 

An icosahedrally symmetric rational map was found in ref.~\cite{HMS}
and this exact same map is reproduced by minimizing over all degree 17
maps.  The value $\ci=363.4$  is given by the $Y_h$ symmetric
buckyball map\footnote{In ref.\cite{HMS} the value of $\ci$ quoted for
the $B=17$ buckyball map was the result of a typographical error and
should read $363.41$,  not $367.41.$} \be
R=\frac{17z^{15}-187z^{10}+119z^5-1}{z^2(z^{15}+119z^{10}+187z^5+17)}\,.
\label{17}
\ee Further confirmation that this is indeed the minimum energy
configuration at this charge comes from non-linear field theory
simulations. We have performed relaxations of initial configurations
of $12+5$, $13+4$ and $5+7+5$ all of which relax very quickly to the
buckyball structure. 

In addition to the buckyball map there is also a non-fullerene map
which has a low value of $\I.$ This map has $O_h$ symmetry, and again
we shall describe its group theory construction by embedding it into
the tetrahedral group. The tetrahedral decomposition in this case
reads \be  \underline{18}\vert_T=3E'+3E_1'+3E_2'\,, \ee with the
two-parameter family of $T$ maps corresponding to the $3E'$  component,
reducing to a one-parameter family of $O$ maps by setting one of the
two parameters to zero. If the remaining parameter is real then the
symmetry extends to $O_h$ and the map is given by $R=p/q$ where \be
p=(129+a)+(2380+116a)z^4+(24310+286a)z^8
+(6188-156a)z^{12}+(17+9a)z^{16}\,,
\label{17b}
\ee and $q(z)=z^{17}p(1/z).$ The minimal $O_h$ map in this class has
$\I=367.2$ when $a=280.9.$ 

Given that Buckyball map is the minimum map with $B=17$ and it is
reproduced in the numerical field theory relaxations, we conclude that
the fullerene hypothesis and the conjecture of ref.~\cite{BS2} are
spectacularly confirmed at this charge; ${\cal S}_{17}$ has $Y_h$
symmetry and the associated polyhedron is the buckyball. 

\subsection{$B=18$}

After the particularly  high symmetry of the $B=17$ solution, the
minimizing map of degree 18 is relatively unremarkable having only
$D_2$ symmetry, and takes the form \be
R=\bigg(\sum_{j=0}^9a_jz^{2j}\bigg)/\bigg(\sum_{j=0}^9a_{9-j}z^{2j}\bigg)\,,
\label{18}\ee
where $a_9=1$ and $a_0,..,a_8$ are complex parameters. The minimum
value is $\I=418.7$ which is attained when the parameters take the
values given in table~\ref{tab-coeff18}. The associated polyhedron,
which is of fullerene type, but is not an isolated pentagon
structure\footnote{In fact, no isolated pentagon structures exist with
12 pentagons and 22 hexagons~\cite{atlas}.}, is difficult to describe
to similar reasons to the $B=14$ and $B=16$ solutions
 with $D_2$ symmetry. It contains 12
pentagons and 22 hexagons, and can best be described, as for $B=16$,
in terms of two half shells which fit together to create the whole
polyhedron. Each shell can be created by first taking a hexagon and
connecting to it, in cyclic order, two hexagons, a pentagon, two
hexagons and another pentagon. Now connect a pentagon, pointing
downward to each of the hexagons, then fill in the gaps, of which
there are 6, with hexagons.

\begin{table}
\centering
\begin{center}
\begin{tabular}{|c|c|c|c|c|c|c|c|c|c|}
\hline & $a_0$ & $a_1$ & $a_2$ & $a_3$ & $a_4$ & $a_5$ & $a_6$ & $a_7$
& $a_8$ \\ \hline  ${\rm Re}(a)$&
-0.1&-5.6&3.2&51.5&-35.9&-50.9&-168.6&-0.3&-10.8\\ \hline  ${\rm
Im}(a)$&-0.5&-16.4&-0.8&104.8&-73.2&10.7&51.4&-3.3&1.4\\ \hline
\end{tabular}
\end{center}
\caption{The coefficients of the minimal $D_2$ map with $B=18$.}
\label{tab-coeff18}
\end{table}

Given the rather low symmetry of the minimum energy rational map, one
might think that as, for example, with the cases of $B=14$ and $B=16$
that there might be some confusion. However, by  relaxing initial
conditions comprizing of $9+9$ and $17+1$ we have reproduced the $D_2$
configuration. Hence, we conclude that ${\cal S}_{18}$ is of the
fullerene type and can be well approximated by the $D_2$ rational map
above.

\subsection{$B=19$}

The minimum value at degree 19 is $\I=467.9$ and is attained by a
$D_3$ symmetric map of the form 
\be
R=\bigg(\sum_{j=0}^6a_jz^{3j+1}\bigg)/\bigg(\sum_{j=0}^6a_{6-j}z^{3j}\bigg)\,,
\label{19a}
\ee
where $a_6=1$ and $a_0,..,a_5$ are complex parameters given in
table~\ref{tab-coeff19}. The associated polyhedron can be constructed
by taking two sets of three hexagons which are almost flat. To each, 
connect a pentagon in the gap connecting two hexagons and another
pentagon next to it connected to only one of the hexagons in the
triple. These two structures are $C_3$ symmetric and can be thought of
as forming the top and bottom of the polyhedron. They are connected
together by 9 sets of two hexagons, one which connects to the top and
the other which connects to the bottom. This is a fullerene type
polyhedron containing 12 pentagons and 24 hexagons.

\begin{table}
\centering
\begin{center}
\begin{tabular}{|c|c|c|c|c|c|c|}
\hline &$a_0$ & $a_1$ & $a_2$ & $a_3$ & $a_4$ & $a_5$ \\ \hline  ${\rm
Re}(a)$&5.2&-0.9&71.4&-325.4&-116.0&0.83\\ \hline  ${\rm
Im}(a)$&0.9&73.6&41.8&-96.7&95.9&-32.5\\ \hline
\end{tabular}
\end{center}
\caption{The coefficients of the  minimal $D_3$ map with $B=19$.}
\label{tab-coeff19}
\end{table}

There is a more symmetric map, with $T_h$ symmetry, whose $\I$ value
is only slightly higher than the minimum at $\I=469.8.$ Computing the
relevant decomposition  \be
\underline{20}\vert_T=4E'+3E_1'+3E_2'\,,\label{decomp20} \ee shows
that there is a three parameter family of tetrahedral maps of degree
19 corresponding to the first component in (\ref{decomp20}).  This
family of maps is given by $R=p/q$ where
\begin{eqnarray}
p= && (239-9b)z+(503a-25c)z^3+(-5508+460b)z^5+(-1300a+284c)z^7
\nonumber\\ && +(-4862-286b)z^9+(1794a+210c)z^{11}+(9996-196b)z^{13}
\nonumber\\ && +(-484a+44c)z^{15}+(135+31b)z^{17}+(-a-c)z^{19}\,,
\label{19b}
\end{eqnarray}
and $q(z)=z^{19}p(1/z).$ The minimum in this family is $T_h$ symmetric
with $a=5.5$, $b=6.3$, $c=37.3$ and produces the $\I$ value given
above. The associated polyhedron contains many four valent bonds and
is most definitely not of the fullerene type. In fact a cursory glance
at the configuration might convince one that the polyhedron has cubic
symmetry with there being eight sets of three fuzed pentagons
effectively situated at the corners of a cube. However, the two
pentagons which are situated at the centre of each face break the
cubic symmetry since they point alternately in different
directions. In total the configuration comprizes of a total of 36
pentagons. 

The $D_3$ configuration was reproduced in the collision and subsequent
relaxation of $17+2$ and therefore we conclude that it corresponds to
${\cal S}_{19}$. It is of the fullerene type and can be
approximated using  the rational map (\ref{19a}).

\subsection{$B=20$}

For $B=20$ the minimum value is $\I=519.6$ and is reproduced by the
$D_{6d}$ symmetric map \be
R=\frac{z^2(ia+bz^6+icz^{12}+z^{18})}{1+icz^6+bz^{12}+iaz^{18}}\,,
\label{20}
\ee with $a=-16.8$,$b=-288.3$ and $c=215.8.$ The associated polyhedron
can be thought of as being created from two half shells connected
together and rotated relative to each other by $30^{\circ}$ to give
$D_{6d}$ symmetry. Each half  can be constructed from a single
hexagon, surrounded by another six forming an almost flat hexagonal
structure, which are then surrounded by 6 hexagons and 6
pentagons. The flat structure has 12 positions for attaching another
polygon, 6 places to connect to one hexagon and 6 places to connect to
two. The pentagons connect to two and the hexagons connect to one
forming a structure which is of the isolated pentagon fullerene type
(corresponding to configuration 72:1 in ref.~\cite{atlas}).
It contains 12 pentagons and 26 hexagons, was reproduced in the
relaxation of $17+3$ and, hence, we conclude that it is ${\cal
S}_{20}$.

\subsection{$B=21$}

The $B=21$ minimizing map is $T_d$ symmetric.
From the decomposition \be \underline{22}\vert_T=3E'+4E_1'+4E_2'
\label{decomp22}
\ee there is a three parameter family
of $T$ symmetric maps corresponding to the $4E_1'$
component and this family of maps is given by $R=p/q$ where
\begin{eqnarray}
p= && 1025+3\,a+b+c+i\left (210+890\,a+74\,b-10\,c \right ) \sqrt
{3}{z}^{2}\nonumber \\ &&+\left (5985+6327\,a+1433\,b-75\,c\right
){z}^{4}  \nonumber \\ && +i \left (54264-7752\,a-680\,b+392\,c\right
)\sqrt {3}{z}^{6} \nonumber \\ &&+\left (
203490+5814\,a-1598\,b+690\,c\right ){z}^{8} \nonumber \\ &&+i\left
(352716+ 16796\,a+2652\,b+260\,c\right )\sqrt {3}{z}^{10} \nonumber \\
&& +\left (293930-25194\, a+442\,b+130\,c\right ){z}^{12} \nonumber \\
&&+i\left (116280-7752\,a-1768\, b-120\,c\right )\sqrt {3}{z}^{14}
\nonumber \\ &&  +\left (20349+14535\,a+221\,b-243\,c \right ){z}^{16}
\nonumber\\ &&+i\left (1330-646\,a+234\,b-10\,c\right ) \sqrt
{3}{z}^{18} \nonumber \\ &&+\left (21+51\,a+13\,b+9\,c\right ){z}^{20}\,,
\label{21}
\end{eqnarray}
and $q(z)=z^{21}p(1/z).$ The minimum of $\I=569.9$ is obtained when
$a=20.8$, $b=-102.0$, and $c=570.1$,
for which the symmetry extends to $T_d$ since $a,b,c$ are all real.
 The associated
polyhedron can be thought of in terms of four copies of two different
pseudo faces, one set is placed on the vertices of a tetrahedron and
the other on the vertices of a tetrahedron dual to the first. In this
respect it is very similar to the $T_d$ configuration with $B=9$, and
different to the $T$ configuration with $B=15$. One set of pseudo
faces comprize of a hexagon triple, whereas the others consists of a
hexagon surrounded alternately by hexagons and pentagons.

Note that this map is the latest in an infinite family of tetrahedral
maps,  corresponding to charges $B=6n+3$, where $n=0,1,2,3,...$. This
is because \be \underline{6n+4}\vert_T=nE'+(n+1)E_1'+(n+1)E_2'
\label{famtet}
\ee so there is an $n$ parameter family of tetrahedral maps
corresponding to the middle component in the above. For $n=0,2,3$
$(B=3,15,21)$ we have seen that this family includes the minimal map,
and for $n=1$ $(B=9)$ this family includes a map which is very close
to the minimal value. Thus it seems possible that other members of
this family will be minimal maps, for example, for $B=27$, although
this configuration must have only $T$ symmetry if it is to be of
the fullerene type and will therefore be similar to that with $B=15$. 

This solution was reproduced in the relaxation of $17+4$ and $20+1$,
and hence we conclude that it corresponds to ${\cal S}_{21}$. It
is of the isolated pentagon fullerene type (corresponding to
configuration 76:2 in ref.~\cite{atlas}), comprizing of 12 pentagons
and 28 hexagons and can be reproduced by the rational map (\ref{21}).

\subsection{$B=22$}

The minimum value at degree 22 is $\I=621.6$, obtained from a $D_{5d}$
symmetric map 
\be
R=\frac{az^2+ibz^7+cz^{12}+idz^{17}+z^{22}}{1+idz^5+cz^{10}+ibz^{15}+az^{20}}
\,,
\ee
where $a=24.8,b=-814.6,c=-2000.3,d=320.3.$

The polyhedron associated with this configuration
 can be constructed in two halves, which fit
together as with many of the solutions already described. To construct
each half, take a pentagon and surround it by 5 hexagons. There are
10 places to position another polygon, 5 of which connect to two
hexagons and 5 which connect to just one. Place 5 pentagons in the
gaps connecting to two hexagons and 5 hexagons just connecting to
one. Then place a further 5 hexagons, connecting to the
pentagons. This configuration, which comprizes of 12 pentagons and 30
hexagons, is of the isolated pentagon fullerene type (it corresponds
to configuration 80:1 in ref.~\cite{atlas}).

For $B=22$ we find that there is an interesting phenomenon in that
an icosahedrally symmetric fullerene polyhedron 
exists~\cite{atlas}, but no corresponding
rational map generated Skyrmion. The $Y_h$ symmetric $C_{80}$ fullerene
is constructed in a similar manner to the $D_{5d}$ fullerene described
above, except that one interchanges the pentagons and hexagons at the
point at which there was a choice in inserting polygons into the 10 
positions. It is easy to check that there are no $Y$ symmetric
rational maps of degree 22, since $\underline{23}\vert_Y$ contains
only representations of dimension three and higher. Thus there are
symmetric fullerene polyhedra which do not correspond to symmetric rational
maps. 

This apparent puzzle can be understood\footnote{We thank Conor Houghton
for pointing this out to us.} by realizing 
that there are rational map generated
Skyrmions whose baryon (and energy) density has more symmetry than
the Skyrme field itself. 
As mentioned earlier, the baryon density of a Skyrmion is localized
around the edges of a polyhedron with the face centres of the
polyhedron  given by the vanishing of the derivative of the rational
map, or more accurately by the zeros of the Wronskian of the numerator
and the denominator \be w(z)=p'(z)q(z)-q'(z)p(z)\,,
\label{wronskian2}
\ee 
which is in general a degree $2B-2$ polynomial in $z.$
For $B=22$ the Wronskian is, therefore, a degree 42 polynomial,
and although there are no $Y$ symmetric degree 22 rational maps
there is a $Y$ symmetric degree 42 polynomial, given by the
product of the Klein polynomials corresponding to the edges and
vertices of an icosahedron~\cite{Kl}. Therefore, it appears that
the existence of a symmetric fullerene polyhedron coincides with the
existence of a rational map whose Wronskian has this symmetry,
but that the existence of such a Wronskian does 
not imply the existence of a symmetric rational map itself.
Fortunately, we have
not encountered this situation in our study of minimal energy
rational maps and Skyrmions for the other charges we have studied,
since it would make the problem of
identifying and constructing a particular rational map a much  more difficult
exercise.

There exists a family of $D_3$ symmetric rational maps with $B=22$ of
the form \be
R=\bigg(\sum_{j=0}^7a_jz^{3j+1}\bigg)/\bigg(\sum_{j=0}^7a_{7-j}z^{3j}\bigg)\,,
\label{22b}
\ee where $a_7=1$ and $a_0,..,a_6$ are complex parameters. The minimum
value of $\ci$ in this class of maps is $\ci=623.4$, which is very
close to that for the $D_{5d}$. The coefficients at this minimum are
presented in table~\ref{tab-coeff22b}. This configuration,  which is
also of the isolated pentagon fullerene type, corresponds to
configuration 80:4 in ref.~\cite{atlas}. It can be constructed by first
taking two hexagon triples each of which are surrounded by 3 pentagons
and 6 hexagons in a $C_3$ arrangement, hexagons filling the gaps which
connect to 2 of the hexagons in the original triples. Then connect 3
more pentagons to each `half' in between each of the $C_3$ symmetric
hexagon triples. The two `halves' should then be connected by a band
of 12 hexagons around the centre, which is split up into 3 lots of 4
by the $C_3$ symmetry. The whole polyhedron contains 12 pentagons and
30 hexagons, as for the $D_{5d}$ configuration.

\begin{table}
\centering
\begin{center}
\begin{tabular}{|c|c|c|c|c|c|c|c|}
\hline & $a_0$ & $a_1$ & $a_2$ & $a_3$  & $a_4$ & $a_5$ & $a_6$ \\
\hline  ${\rm Re}(a)$&4.5&-75.4&-393.4&270.5&26.1&123.8&41.5 \\ \hline
${\rm Im}(a)$&-3.2&54.9&62.3&391.5&872.7&-177.2&13.8 \\ \hline
\end{tabular}
\end{center}
\caption{The coefficients of the  minimal $D_3$ map with $B=22$.}
\label{tab-coeff22b}
\end{table}

Relaxation of clusters containing $17+5$, $20+2$ and $21+1$ all lead
to the same structure, that with $D_3$ symmetry. Therefore, we
conclude that ${\cal S}_{22}$ is that approximated by the rational map
(\ref{22b}). This removes, from the point of view of this paper at
least, further motivation for attempting to create the Skyrmion with
the $Y$ symmetric baryon density isosurface.

\subsection{Summary}

The main conclusion of this section on Skyrmion identification is that
the fullerene hypothesis appears to apply for a wide range of charges,
and that the rational map ansatz can be used to make a good
approximation to the solutions. In particular, we have
concluded\footnote{We should note that we have also confirmed the
results of refs.~\cite{BS2,HMS} for $1\le B \le 8$.} that ${\cal S}_B$
is of the fullerene type for $7\le B \le 22$, except when $B=9$ and
$B=13$. For these charges  the associated polyhedron contains
four-valent bonds, but as we shall discuss in
section~\ref{sec-enhance}, even these solutions can be related to
fullerene polyhedra via symmetry enhancement. Clearly, there is a
strong correlation between the structure of
multi-Skyrmions and that of fullerene polyhedra.

For $B=17$, $B=20$, $B=21$ and $B=22$ where there are
fullerene polyhedra in which all the pentagons are surrounded by just
hexagons, this type of configuration is picked out as that with
minimum energy. In fullerene chemistry the isolated pentagon isomers
are thought to minimize energy by placing the pentagonal defects as far
as possible from each other, and it likely that this also taking place
here.
It is interesting to speculate that these highly
spherical, isolated pentagon configurations will be the minima at
higher charge in a similar way to our suggestion of the fullerene
hypothesis just on the basis of the $B=7$, $B=8$ minima and the
$B=9$ saddle point. Although there will no doubt be caveats, as we
have reported in this paper for the fullerene hypothesis for $B=9$ and
$13$, we believe
they are likely to be the exception rather than the rule.

Notwithstanding these successes we have turned up a few
oddities. Firstly, we have seen that in the cases of $B=10$, $B=14$,
$B=16$ and $B=22$, that the minimum energy rational maps might not be
${\cal S}_B$. For $B=10$ the situation is particularly interesting
since the minimum energy rational map is not of the fullerene type,
yet only fullerenes are produced in the relaxation of initially
well-separated clusters. While for $B=14$ we have been unable to
produce a rational map which accurately reproduces the configuration
which is readily produced in the full relaxation, probably since it is
elongated and has little symmetry. The cases of $B=16$ and $B=22$ are
probably more mundane since the differences in the $\ci$ values for
the two fullerene-like configurations are very small, and it is not
difficult to imagine that using $\ci$ as the energy functional rather
than the true energy manages to swap around the two
configurations. These special cases vindicate our approach of using
two different methods to make our identifications.

\section{Skyrmion energies}\news
\label{sec-energies}

\begin{table}
\centering
\begin{tabular}{|c|c|c|c|c|c|c|c|}
\hline $B$ & $G$ & $E_{\rm dis}$ & $B_{\rm dis}$ & $E/B$ & $E_{\rm
dis}$ & $B_{\rm dis}$ & $E/B$ \\ \hline 

    1 & $O(3)$ &1.1591 &   0.9407 &   1.2322 &   1.2137 &   0.9849 &
    1.2322 \\ 2 & $D_{\infty h}$ &   2.2335 &   1.8935 &   1.1796 &   2.3260 &
    1.9726 &   1.1791 \\ 3 & $T_d$ &   3.2773 &   2.8573 &   1.1470 &
    3.3960 &   2.9627 &   1.1462 \\ 4 & $O_{h}$ &   4.2683 &   3.8091 &
    1.1205 &   4.4265 &   3.9519 &   1.1201 \\ 5 & $D_{2d}$ &   5.3308 &
    4.7708 &   1.1174 &   5.5199 &   4.9409 &   1.1172 \\ 6 & $D_{4d}$ &
    6.3391 &   5.7230 &   1.1077 &   6.5692 &   5.9296 &   1.1079 \\ 7
    & $Y_h$ &   7.3243 &   6.6889 &   1.0950 &   7.5766 &   6.9210 &
    1.0947 \\ 8 & $D_{6d}$ &   8.3796 &   7.6441 &   1.0962 &   8.6690 &
    7.9100 &   1.0960 \\ 9 & $D_{4d}$ &   9.4026 &   8.5984 &   1.0936 &
    9.7322 &   8.8990 &   1.0936 \\ 10 & $D_{4d}$ &  10.4212 &   9.5579 &
    1.0903 &  10.7826 &   9.8893 &   1.0903 \\ 11 & $D_{3h}$ &  11.4464 &
    10.5129 &   1.0888 &  11.8457 &  10.8788 &   1.0889 \\ 12 & $T_d$ &
    12.4533 &  11.4721 &   1.0855 &  12.8888 &  11.8723 &   1.0856 \\
    13 & $O$ &  13.4689 &  12.4304 &   1.0835 &  13.9311 &  12.8585 &
    1.0834 \\ 14 & $D_{2}$ &  14.5057 &  13.3819 &   1.0840 &  15.0139 &
    13.8480 &   1.0842 \\ 15 & $T$ &  15.5214 &  14.3403 &   1.0824 &
    16.0635 &  14.8387 &   1.0825 \\ 16 & $D_3$ &  16.5274 &  15.2969 &
    1.0804 &  17.0167 &  15.8283 &   1.0808 \\ 17 &  $Y_h$ &  17.5275 &
    16.2677 &   1.0774 &  18.1205 &  16.8185 &   1.0774 \\ 18 & $D_2$ &
    18.5677 &  17.2152 &   1.0786 &  19.2134 &  17.8094 &   1.0788 \\
    19 & $D_3$ &  19.6234 &  18.1913 &   1.0787 &  20.2717 &  18.7951 &
    1.0786 \\ 20 & $D_{6d}$ &  20.6414 &  19.1607 &   1.0773 &  21.3198 &
    19.7781 &   1.0779 \\ 21 & $T_d$ &  21.7056 &  20.1351 &   1.0780 &
    22.3781 &  20.7580 &   1.0780 \\ 22 & $D_{5d}$ &  22.7349 &  21.1146 &
    1.0767 &  23.4183 &  21.7525 &   1.0766 \\

\hline
\end{tabular}
\caption{The results of relaxing the rational map solutions which
minimize the function $\ci$. The first set correspond to $N=100$ and
$\Delta x=0.02$, while the second set have $N=200$ and $\Delta
x=0.01$. In both cases the profile function was set to be zero at
$r_{\infty}=10$. Notice that the $E/B$ values are very close for the
two different size grids suggesting that this figure is universal with
an error $\pm 0.001$.}
\label{tab-relax1}
\end{table}

\begin{table}
\centering
\begin{tabular}{|c|c|c|c|c|c|c|c|}
\hline $B$ & $G$ & $E_{\rm dis}$ & $B_{\rm dis}$ & $E/B$ & $E_{\rm
dis}$ & $B_{\rm dis}$ & $E/B$ \\ \hline 

    9* & $T_d$ &   9.4281 &   8.5935 &   1.0971 &   9.7636 &   8.8975 &
   1.0973 \\ 10* & $D_3$ &  10.4223 &   9.5592 &   1.0903 &  10.7826 &
   9.8891 &   1.0904 \\ 10* & $D_{3d}$  &  10.4171 &   9.5586 &   1.0898 &
   10.7791 &   9.8890 &   1.0900 \\ 10* & $D_{3h}$ &  10.4623 &   9.5548 &
   1.0950 &  10.8308 &   9.8882 &   1.0953 \\ 13* & $D_{4d}$ &  13.4674 &
   12.4295 &   1.0835 &  13.9311 &  12.8585 &   1.0834 \\ 13* & $O_h$ &
   13.6152 &  12.4264 &   1.0957 &  14.0939 &  12.8548 &   1.0964 \\
   15* & $T_d$ &  15.5792 &  14.3429 &   1.0862 &  16.1226 &  14.8369 &
   1.0867 \\ 16* & $D_{2}$ &  16.5294 &  15.2964 &   1.0806 &  17.1092 &
   15.8281 &   1.0809 \\ 17* & $O_h$ &  17.5351 &  16.2654 &   1.0781 &
   18.1389 &  16.8215 &   1.0783 \\ 19* & $T_h$ &  19.6461 &  18.1854 &
   1.0803 &  20.2996 &  18.7935 &   1.0801 \\ 22* & $D_3$ &  22.7455 &
   21.1162 &   1.0772 &  23.4175 &  21.7509 &   1.0766 \\

\hline
\end{tabular}
\caption{Same as table~\ref{tab-relax1} but for the other saddle
points of $\ci$ listed in table~\ref{tab-sa2}.}
\label{tab-relax2}
\end{table}

It is usual when presenting numerically generated minimum energy
Skyrmion configurations to discuss their energy. Before we present
what we believe are good representations of the energies, we should
just make a point that identifying the symmetry is probably a better
way of judging success, rather than just this single number. In particular,
making comparisons between different approaches for computing the
energy presented in the literature is extremely hazardous, whereas the
symmetry identification should be universal.

The approach that we shall discuss here is based on using the full
non-linear dynamics to relax a solution generated using the rational
map ansatz for a given charge and symmetry. We have already seen
that the rational map ansatz systematically over estimates the
ratio of the energy to baryon number ($E/B)$ for a given configuration
and it is clear that the numerical relaxation is likely to reduce this
somewhat. 

Specifically, we have performed relaxations on numerical grids for all
the solutions listed in table~\ref{tab-sa1} and table~\ref{tab-sa2}
with (a) $\Delta x=0.02$ and $N=100$ and (b) $\Delta x=0.01$ and
$N=200$, both using Dirichlet (fixed) boundary conditions at the edge
of the grid. These two discretized grids have exactly the same spatial
extent and so when computing the initial profile function for the
rational map ansatz we have set the profile function to zero at
$r_{\infty}=10$. The results of this extensive set of simulations are
presented in table~\ref{tab-relax1} for the solutions which are the
minimum energy rational maps with respect to the energy functional
$\ci$ and in table~\ref{tab-relax2} for the other critical points
of $\ci$, some of which we have concluded in the previous section are,
in fact, the minimum energy Skyrmions. The total simulation time ---
which is the number of timesteps multiplied by $\Delta t$ --- for
$N=100$ is twice that for $N=200$, although making an exact comparison
can be  misleading since the rate at which the relaxation takes place
is somewhat arbitrary due to the periodic removal of kinetic
energy. Suffice to say, in both cases we believe that we have run the
code for long enough for it to settle down to the minimum.

The first thing to notice is that the computed values of the baryon
number for the discrete grid $B_{\rm dis}$ are less than the relevant
integer; the value for $N=200$ being closer than that for
$N=100$. There are two possible sources for this error: the first is
numerical discretization error from the computation of the spatial
derivatives (and the resulting numerical integration)  and the other
is that the physical size of the box $r_{\infty}$ is not large enough
to encompass the whole solution and we have underestimated the
gradient energy. 

In order to understand which is the dominant effect
we have repeated the same relaxations with $r_{\infty}=20$ ($\Delta
x=0.02$ and $N=200$) for $B=1-4,9,13,17$ and the results are presented
in table~\ref{tab-largeen} for a total simulation time which is
exactly the same as that for $\Delta x=0.02$, $N=100$ and
$r_{\infty}=10$. The values of $B_{\rm dis}$ agree to the
third decimal place for $B=1-4$ with the agreement being less good
for larger values of $B$. Clearly discretization error is
the dominant effect for the small values of $B$, and the truncation
error due to $r_{\infty}$ being finite becomes important as it increases.

\begin{table}
\centering
\begin{tabular}{|c|c|c|c|c|}
\hline $B$ & $G$ & $E_{\rm dis}$ & $B_{\rm dis}$ & $E/B$ \\ \hline 

    1 & $O(3)$ &   1.1587 &   0.9410 &   1.2314 \\ 2 &$D_{\infty h}$
 &   2.2324 &   1.8936
    &   1.1789 \\ 3 & $T_d$ &   3.2748 &   2.8562 &   1.1466 \\ 4 & $O_h$ &
    4.2683 &   3.8095 &   1.1204 \\ 9 & $D_{4d}$ &   9.3906 &   8.5870 &
    1.0936 \\ 13 & $O$ &  13.4572 &  12.4197 &   1.0835 \\ 17 & $Y_h$&  17.5511
    &  16.2862 &   1.0777 \\
   
\hline
\end{tabular}
\caption{The results of relaxing the rational map solutions for
selected charges using $N=200$, $\Delta x=0.02$ and
$r_{\infty}=20$. Notice that the numerical values are almost identical
to those listed in table~\ref{tab-relax1} for the same value of
$\Delta x$.}
\label{tab-largeen}
\end{table}

Remarkably, the value of $E_{\rm dis}/B_{\rm dis}$ appears to remain
constant at the level of around $\pm 0.001$ for all the relaxations,  
irrespective of what is the main source of the error in computing the
individual values for  the sensible parameters $\Delta x$
and $N$ that we have used. Therefore, 
the main conclusion we will draw from these simulations is the value
of $E_{\rm dis}/B_{\rm dis}$ which we will equate with the true value
of $E/B$. Clearly, knowledge of $E/B$ to a certain level
of accuracy allows one to compute the energy of the solution, $E_B$, to the
same level of accuracy, and the values of the energies based upon this
hypothesis are presented in table~\ref{tab-compen1} and
table~\ref{tab-compen2} for the minima of $\ci$ and the other critical
points respectively. For the subsequent discussions we will take the
value of $E/B$ to be that  computed when $\Delta x=0.01$, $N=200$ and
$r_{\infty}=10$ subject to an assumed error of $\pm 0.001$, with the
relative difference between adjacent values of $B$ probably being even
more accurate. This error
budget is used to include the many other possible systematic
uncertainties in computing $E/B$ which have not already been discussed
and the spread of the values computed.

The precise values of $E/B$ we have computed here differ from the
those quoted for $B=1-9$ in our earlier work~\cite{BS2}\footnote{Note
that the $B=9$ solution of ref.~\cite{BS2} had $T_d$ symmetry and so
one should be careful to make the correct comparison}, and for $B=1-4$ in
recent work which used a simulated annealing algorithm on the full
field dynamics~\cite{HSW}. By making comparison with ref.~\cite{BS2} at the
level of accuracy suggested there ($\sim 1\%$), we see that only the
$B=2$ and $B=3$ are discrepant and then the difference is only of the
order of an extra $1\%$, but that the actual values quoted are very different 
at the level of 3 decimal places. We believe
the earlier method has a tendency to under estimate the true value of
$E/B$ since the value of the field at the boundary, which is kept
fixed during the relaxation runs, is sensitive to the initial
conditions, which are well-separated in contrast to the situation here.
 This is borne out on inspection of the $E/B$ values computed for
the relaxed solutions from well-separated clusters for large $B$ used to 
identify the minima. The
comparison with the results of ref.~\cite{HSW} is less good, their
results being systematically about $1\%$ higher than those presented
here and $2\%$ higher than those of ref.~\cite{BS2}.
Although it is difficult to make strong conclusions, we believe
that these higher values could well be a function of using periodic
boundary conditions rather than fixed. This assumption was used to
make their computed values of $B$ almost exactly an integer, but it is
clear that such an assumption will modify the scale of the solution. 
In particular, their quoted value for the energy of a single Skyrmion
is larger than the known value $(E=1.232)$, and the solution assuming
periodic boundary conditions with a domain of finite extent will
be somewhat different.

\begin{table}
\centering
\begin{tabular}{|c|c|c|c|c|c|}
\hline $B$ & $G$ & $E/B$ & $E_B$ & $E/B$ & $E_B$ \\ \hline  

    1 & $O(3)$ &   1.2322 &   1.2322 &   1.2322 &   1.2322 \\ 2 &
    $D_{\infty h}$ &   1.1796 &   2.3592 &   1.1791 &   2.3582 \\ 3 &
    $T_d$ &   1.1470 &   3.4410 & 1.1462 &   3.4386 \\ 4 & $O_h$ &
    1.1205 &   4.4820 &   1.1201 & 4.4804 \\ 5 & $D_{2d}$ &   1.1174 &
    5.5870 &   1.1172 &   5.5860 \\ 6 & $D_{4d}$ & 1.1077 &   6.6462 &
    1.1079 &   6.6474 \\ 7 & $Y_h$ &   1.0950 & 7.6650 &   1.0947 &
    7.6629 \\ 8 & $D_{6d}$ &   1.0962 &   8.7696 & 1.0960 &   8.7680
    \\ 9 & $D_{4d}$ &   1.0936 &   9.8424 &   1.0936 & 9.8424 \\ 10 &
    $D_{4d}$ &   1.0903 &  10.9030 &   1.0903 &  10.9030 \\ 11 &
    $D_{3h}$ &   1.0888 &  11.9768 &   1.0889 &  11.9779 \\ 12 & $T_d$
    &   1.0855 & 13.0260 &   1.0856 &  13.0272 \\ 13 & $O$ &   1.0835
    &  14.0855 & 1.0834 &  14.0842 \\ 14 & $D_2$ &   1.0840 &  15.1760
    &   1.0842 & 15.1788 \\ 15 & $T$ &   1.0824 &  16.2360 &   1.0825
    &  16.2375 \\ 16 & $D_3$ &   1.0804 &  17.2864 &   1.0808 &
    17.2928 \\ 17 & $Y_h$ &   1.0774 & 18.3158 &   1.0774 &  18.3158
    \\ 18 & $D_2$ &   1.0786 &  19.4148 & 1.0788 &  19.4184 \\ 19 &
    $D_3$ &   1.0787 &  20.4953 &   1.0786 & 20.4934 \\ 20 & $D_{6d}$
    &   1.0773 &  21.5460 &   1.0779 &  21.5580 \\ 21 & $T_d$ &
    1.0780 &  22.6380 &   1.0780 &  22.6380 \\ 22 & $D_{5d}$& 1.0767 &
    23.6874 &   1.0766 &  23.6852 \\
 
\hline
\end{tabular}
\caption{The actual computed energies, $E_B$, of the solutions presented in
table~\ref{tab-relax1} deduced from $E_B=B*E_{\rm dis}/B_{\rm dis}$.}
\label{tab-compen1}
\end{table}

\begin{table}
\centering
\begin{tabular}{|c|c|c|c|c|c|}
\hline $B$ & $G$ & $E/B$ & $E_B$ & $E/B$ & $E_B$ \\ \hline  

    9* & $T_d$  &   1.0971 &   9.8739 &   1.0973 &   9.8757 \\ 10* &
   $D_3$ &   1.0903 &  10.9030 &   1.0904 &  10.9040 \\ 10* & $D_{3d}$
   &   1.0898 &  10.8980 & 1.0900 &  10.9000 \\ 10* & $D_{3h}$ &
   1.0950 &  10.9500 &   1.0953 & 10.9530 \\ 13* &  $D_{4d}$&   1.0835
   &  14.0855 &   1.0834 & 14.0842 \\ 13* & $O_h$ &   1.0957 &
   14.2441 &   1.0964 &  14.2532 \\ 15* & $T$  &   1.0862 & 16.2930 &
   1.0867 &  16.3005 \\ 16* & $D_2$ &   1.0806 &  17.2896 & 1.0809 &
   17.2944 \\ 17* & $O_h$ &   1.0781 &  18.3277 &   1.0783 & 18.3311
   \\ 19* & $T_h$ &   1.0803 &  20.5257 &   1.0801 &  20.5219 \\ 22* &
   $D_3$ &   1.0772 &  23.6984 &   1.0766 &  23.6852 \\
 
\hline
\end{tabular}
\caption{The actual computed energies, $E_B$, of the solutions presented in
table~\ref{tab-relax2} deduced from $E_B=B*E_{\rm dis}/B_{\rm dis}$.}
\label{tab-compen2}
\end{table}

We should comment on the computed values of $E/B$ for values of $B$
where we have more than one rational map which is of low energy with
respect to $\ci$. Previously in a number of cases ($B=10,16$ and $22$)
we had concluded that the minimum energy rational map is not
necessarily the minimum energy Skyrmion and one might hope that the
relaxation process might confirm this with their energies being
separated by a significant amount.

For $B=9,15,17$ and $19$ its clear that the computed values of $E/B$
are systematically much higher for the solutions which are not minima
of the energy functional $\ci$ and therefore we conclude that these
solutions are definitely not the minima of the full Skyrme energy
functional (remember that on the basis of relaxation of well-separated
clusters we have concluded that the minima with respect to $\ci$ are
in fact the minima). For $B=13$ the solution with $D_{4d}$ symmetry
has exactly the same value as that with $O$, while that with $O_h$ is
much higher. It might seem remarkable that the values for the $D_{4d}$
and $O$ symmetric solutions are exactly the same, but one should
remember that the two solutions are very similar and can be related by
symmetry enhancement (see section~\ref{sec-enhance}). Since the $O$
solution was produced from the relaxation of clusters and the $\ci$
values for the two solutions are very close anyway, we conclude
that the $O$ solution is probably the minimum.  For $B=10,16$ and
$22$, we are unable to tell the different candidate minima  apart based on the
computed energies since our quoted error of $\pm 0.001$ for each of the values
of $E/B$ encompasses the different solutions under consideration. For
$B=10$ the $D_{3h}$ solution is of higher energy and is clearly not
the minima, but the other 3 are
well within the range of uncertainty. This is also the case for $B=16$
and $B=22$, where each of the candidate minima are within the quoted
range for $E/B$. 

\begin{table}
\centering
\begin{tabular}{|c|c|c|c|c|c|}
\hline $B$ & $G$ & $E/B$ & $E_B$ & $I_B$ & $\Delta E/B$ \\ \hline  

    1 & $O(3)$ &   1.2322 &   1.2322 &   0.0000 &   0.0000 \\
    2 & $D_{\infty h}$ &   1.1791 &   2.3582 &   0.1062 &   0.0531 \\
    3 & $T_d$ &   1.1462 &   3.4386 &   0.1518 &   0.0860 \\
    4 & $O_h$ &   1.1201 &   4.4804 &   0.1904 &   0.1121 \\
    5 & $D_{2d}$ &   1.1172 &   5.5860 &   0.1266 &   0.1150 \\
    6 & $D_{4d}$ &   1.1079 &   6.6474 &   0.1708 &   0.1243 \\
    7 & $Y_h$ &   1.0947 &   7.6629 &   0.2167 &   0.1375 \\
    8 & $D_{6d}$ &   1.0960 &   8.7680 &   0.1271 &   0.1362 \\
    9 & $D_{4d}$ &   1.0936 &   9.8424 &   0.1578 &   0.1386 \\
   10* & $D_3$ &   1.0904 &  10.9040 &   0.1706 &   0.1418 \\
   11 & $D_{3h}$ &   1.0889 &  11.9779 &   0.1583 &   0.1433 \\
   12 & $T_d$ &   1.0856 &  13.0272 &   0.1829 &   0.1466 \\
   13 & $O$ &   1.0834 &  14.0842 &   0.1752 &   0.1488 \\
   14** & $C_2$ &   1.0842 &  15.1788 &   0.1376 &   0.1480 \\
   15 & $T$ &   1.0825 &  16.2375 &   0.1735 &   0.1497 \\
   16* & $D_2$ &   1.0809 &  17.2944 &   0.1753 &   0.1513 \\
   17 & $Y_h$ &   1.0774 &  18.3158 &   0.2108 &   0.1548 \\
   18 & $D_2$ &   1.0788 &  19.4184 &   0.1296 &   0.1534 \\
   19 & $D_3$ &   1.0786 &  20.4934 &   0.1572 &   0.1536 \\
   20 & $D_{6d}$ &   1.0779 &  21.5580 &   0.1676 &   0.1543 \\
   21 & $T_d$ &   1.0780 &  22.6380 &   0.1522 &   0.1542 \\
   22* & $D_3$ &   1.0766 &  23.6852 &   0.1850 &   0.1556 \\

\hline
\end{tabular}
\caption{A summary of the symmetry and energy of the Skyrmion
configurations which we have identified as the minima. 
Included also is the ionization energy ($I_B$) --- that required to
remove one Skyrmion --- and the binding energy per Skyrmion ($\Delta E/B$) 
--- that is the energy required to
split the charge $B$ Skyrmion into $B$ with charge one divided by the
total number of Skyrmions. (*) These correspond
to the minimum energy Skyrme solutions which are not minimum 
energy solutions within the rational map ansatz. (**) The
values quoted for $B=14$ are computed using the initial configuration
with $D_2$ symmetry since we have been unable to derive the rational map with $C_2$ symmetry}
\label{tab-ionbind}
\end{table}

\begin{figure}
\centerline{\epsfxsize=10cm\epsfysize=10cm\epsffile{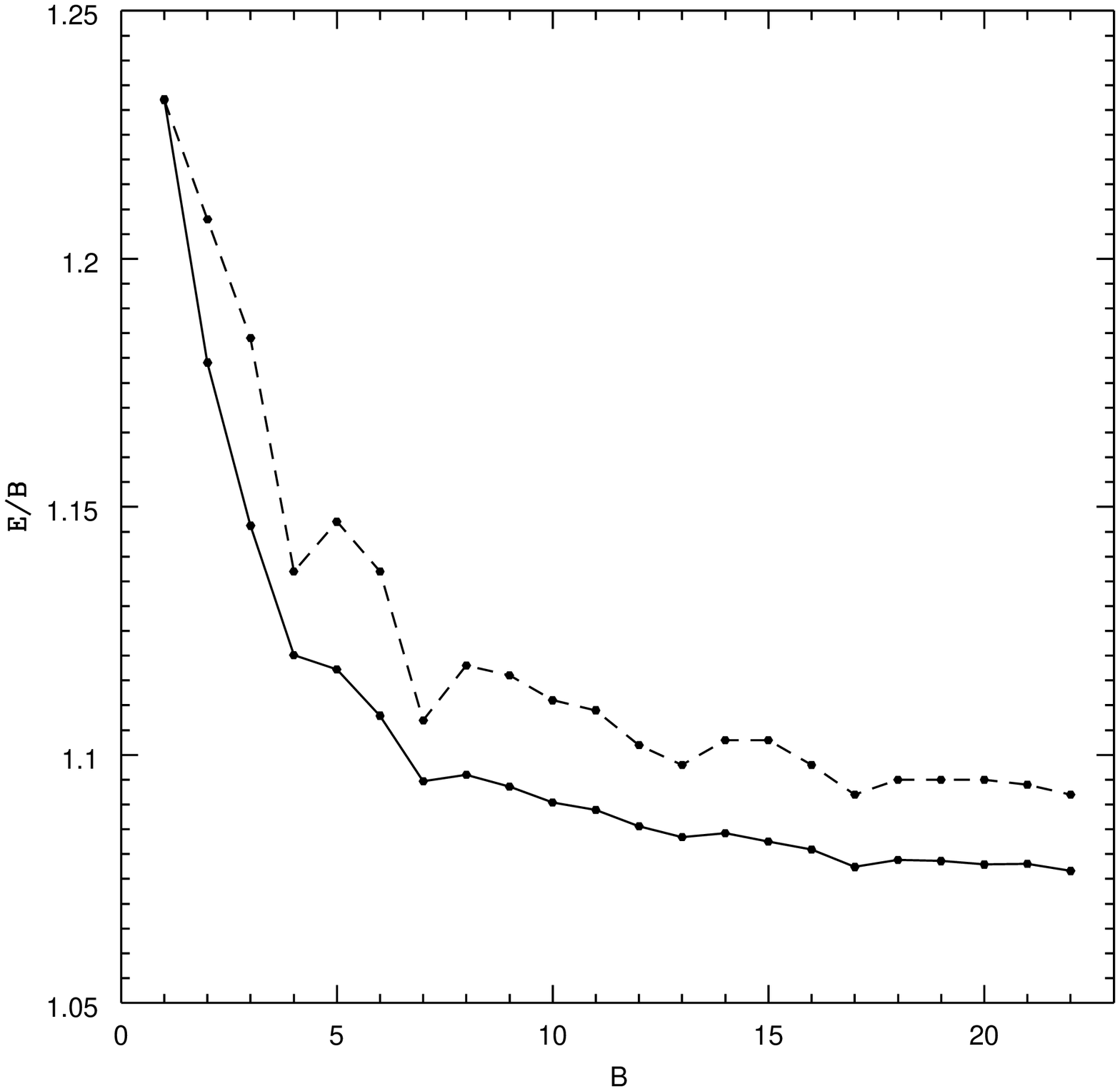}}
\caption{The computed values of $E/B$ as a function of $B$ for the
configurations which we have identified as the minimum energy
solutions, that is, those summarized in table~\ref{tab-ionbind}. The
solid line is that after the process of relaxation, and the dashed
line that from before, that is, the value for the appropriate rational
map. For $B=14$ where we have no rational map to represent the minimum
energy solution we have used the values for the solution with $D_2$
symmetry.}    
\label{fig-eb}
\end{figure}

\begin{figure}
\centerline{\epsfxsize=10cm\epsfysize=10cm\epsffile{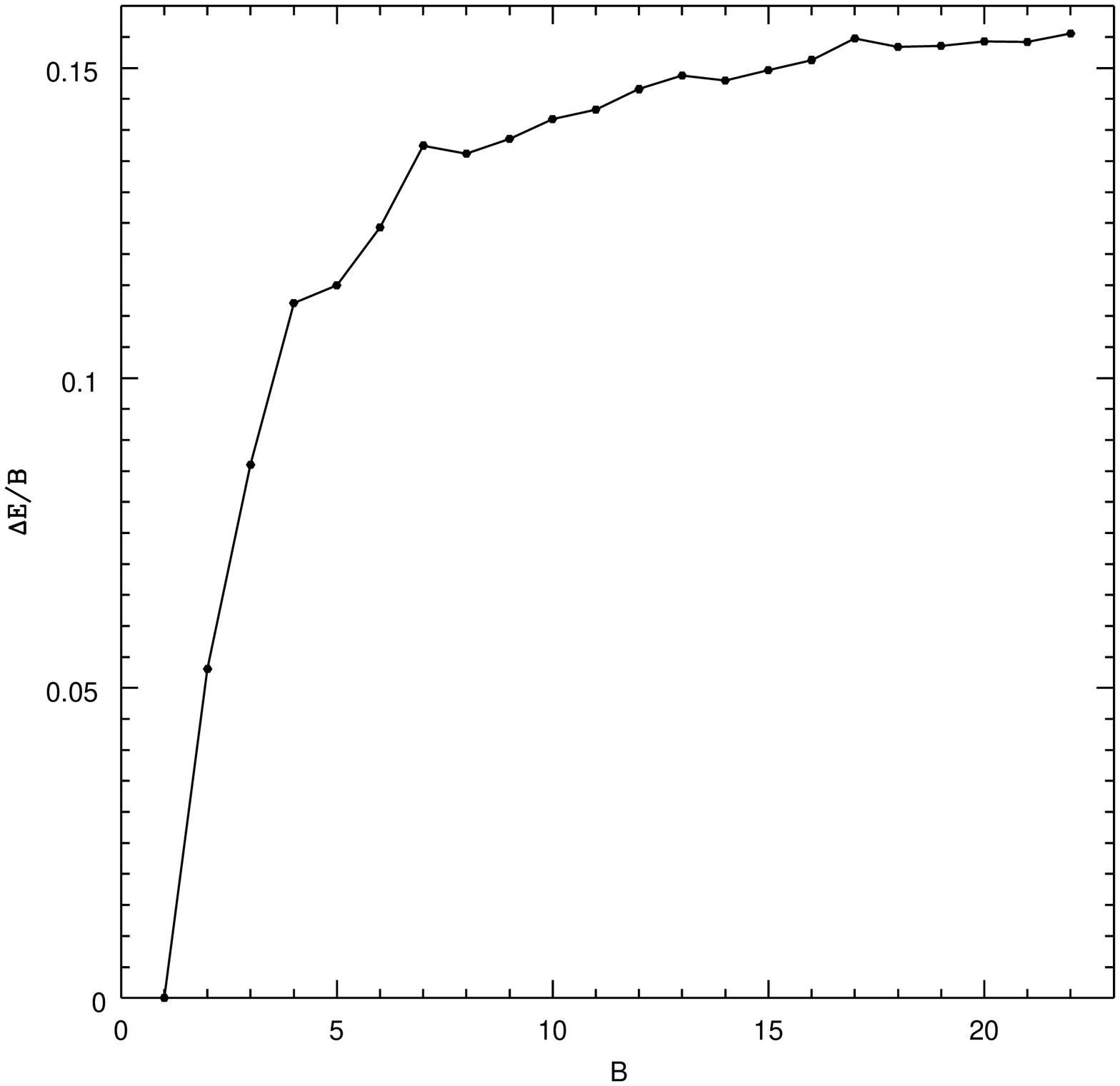}}
\caption{As for fig.~\ref{fig-eb}, but $\Delta E/B$ is plotted instead
of $E/B$. For large $B$ the values appear to level out at around $0.15-0.16$
 as one might expect in a simple model of nuclei excluding the effect of the Coulomb interaction.}    
\label{fig-bind}
\end{figure}

In table~\ref{tab-ionbind} we have summarized the computed values of $E/B$
and $E_B$ for the solutions which we have identified as the minima in
section~\ref{sec-identify}. Included also are the ionization energy
$I_B=E_{B-1}+E_1-E_{B}$, the  energy required to remove a single
Skyrmion, and the binding energy per baryon given by $\Delta
E/B=E_1-(E/B)$, which is the energy required to separate the solution
up into single Skyrmions divided by the total baryon number. The
accuracy of $\Delta E/B$ will be exactly that of the computed value of
$E/B$ since the value of $E_1$ we compute by this method appears to
be exact within the quoted limits, but the errors in computing $I_B$ 
could theoretically be
larger since it is the difference of two energies.
 For $B>10$ the worst case errors in computing $I_B$ could be
as much as $\pm 0.02$ (a significant amount on inspection of the
quoted values), but since we have already commented that we
believe the  difference in energies for adjacent values of $B$ will be
even more accurate than the absolute errors in the energy we suspect
that things will be much better. We will comment on this in subsequent
sections. 

The values of $E/B$ computed for these relaxed solutions and also for the
original rational map are plotted against $B$ in
fig.~\ref{fig-eb}. Both start at approximately the same value for $B=1$,
and both appear to asymptote for large $B$, albeit at different
values. For the relaxed solutions this appears to be about $6-7\%$ above
the Faddeev-Bogomolny bound, which is compatible with that computed
for the hexagonal Skyrme lattice~\cite{BS3}, which can be thought of
as the infinite limit of a shell-like Skyrmion, while the asymptote for
the rational map ansatz is higher at around $9\%$. The curve for the
relaxed solutions is smoother than that for the rational maps,
which has notable dips associated with the highly symmetric solutions
with $B=4,7,13$ and $17$. Although these deviations from what appears to
be approximately a smooth curve do not totally disappear after
relaxation, one can deduce that the other solutions, not being
particularly spherical, do not fit the rational map ansatz as well, but
that the relaxation using the full non-linear field dynamics softens these effects. 

The binding energy per baryon, $\Delta E/B$, is plotted against $B$ in
fig.~\ref{fig-bind}, its shape just being the inversion of
fig.~\ref{fig-eb}. Interestingly, it increases to an asymptote just as
one might expect in a simple model of nuclei which excludes the
Coulomb interaction within the nucleus. We shall return to this issue
in section~\ref{sec-app}.
 
\begin{figure}
\centerline{\epsfxsize=10cm\epsfysize=10cm\epsffile{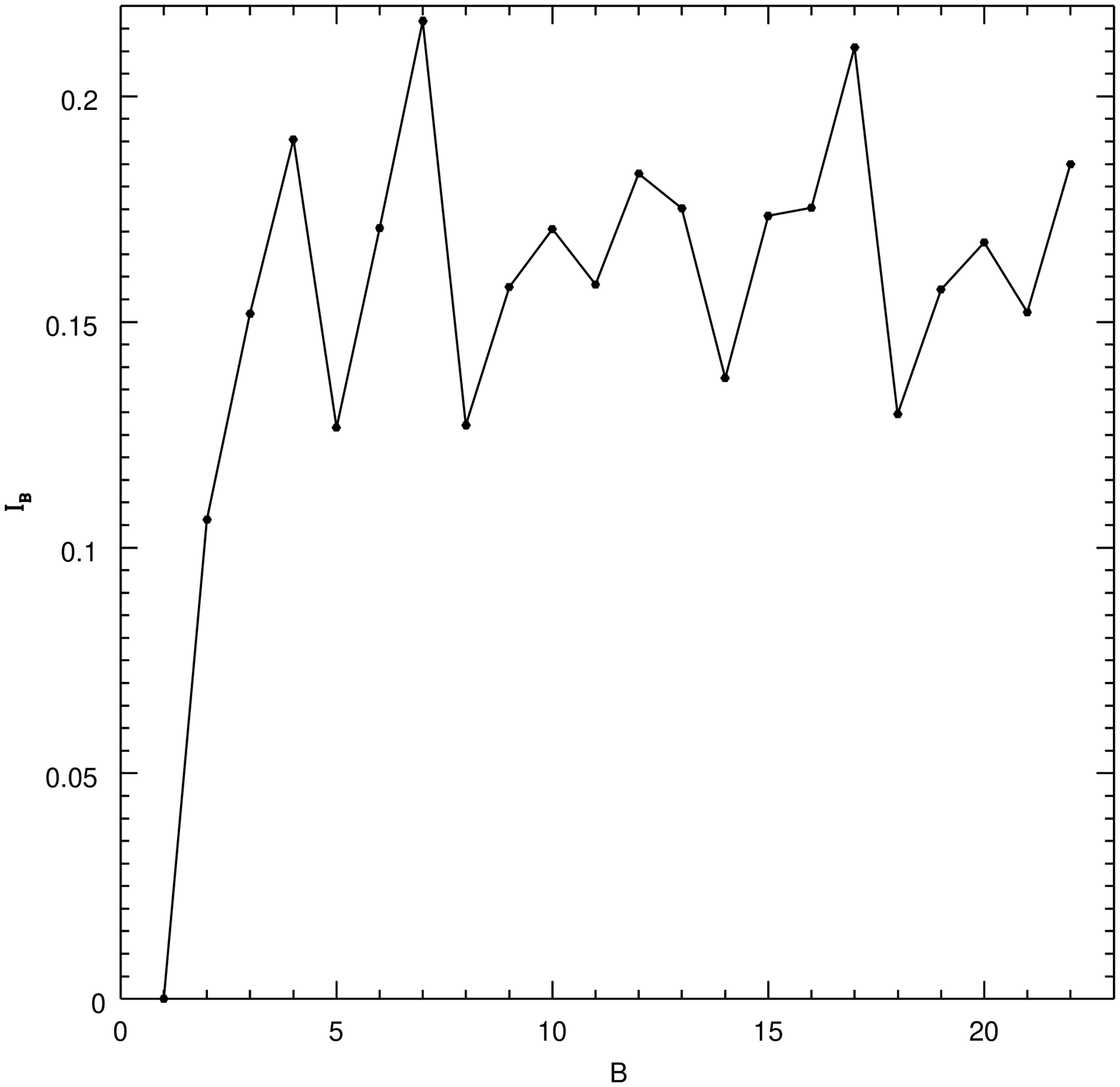}}
\caption{As for fig.~\ref{fig-eb}, but $I_B$ is plotted instead
of $E/B$. Notice that the most stable solutions are those with the most symmetry, $B=4,7$ and 17, while the least stable are those with little symmetry $B=5,8,14$ and $18$.}    
\label{fig-i}
\end{figure}

The ionization energy, $I_B$, is plotted against $B$ in
fig.~\ref{fig-i}. We have already commented that our quoted errors in
$E/B$ might lead to substantial errors in computing $I_B$, but that
systematic errors in computing the energy of a particular solution are
likely to be similar and therefore our computed values for $I_B$ 
could probably be more accurate than one might naively
expect. This is borne out by a cursory inspection of fig.~\ref{fig-i}:
we see that the most stable solutions with respect to the removal of a
single Skyrmion are those with the most symmetry $B=4,7$ and $17$,
while those with the least symmetry $B=5,8,14$ and $18$ are much less
stable. This is very much as one might expect.

\section{Discussion}\news

There are some remarkable aspects of the solutions which we have
created. In this section we point out, discuss and attempt to explain some of
them.

\subsection{Platonic symmetries}
\label{sec-plat}

It had been known for sometime that Skyrmion solutions existed whose
associated polyhedra are platonic solids ($B=3,4$ and 7) and it had
been conjectured that the minimum energy solution with $B=17$ had
$Y_h$ symmetry. These platonic symmetry groups are the groups with the highest
symmetry (most generators) which are discrete subgroups of $O(3)$, the dihedral
groups having considerably fewer generators.
Here, we have shown that platonic symmetries are even
more prevalent in Skyrmion solutions, be they minimum energy solutions
$B=12,13,15$ and $21$, or low energy saddle points $B=9,13,15,17$ and
$19$.  The existence of these solutions is truly remarkable.

This could have been expected within the fullerene hypothesis since the
platonic groups $T$ and $Y$ are compatible with the associated
polyhedron comprizing of pentagons and hexagons. But we have also seen
that the existence of a highly symmetric fullerene polyhedron
compatible with the fullerene hypothesis at a particular charge does
not necessarily imply that it is a minimum energy configuration. In
particular, tetrahedral fullerene structures are compatible with
$B=16$ and $19$, and the group theory decomposition is consistent
with the existence of appropriate rational maps.
However, in each of these cases the
minimum energy Skyrmion has much less symmetry. Clearly, the solution
being highly symmetric is not the sole criterion in minimizing the
Skyrme energy functional. We have also encountered one case, 
icosahedral symmetry for $B=22$, in which a platonic fullerene
polyhedron exists,
but no corresponding platonic rational map (and probably no
Skyrmion either). However, it appears that a rational map exists
(and hence a corresponding Skyrmion) for which the baryon density
surface has more symmetry than the Skyrme field and is icosahedrally symmetric.
But again this very symmetric structure appears not to be
the minimum energy Skyrmion.

For $B>7$ the octahedral group is incompatible with the fullerene
hypothesis since it requires four-fold symmetry which is impossible
in a polyhedron comprizing only of pentagons and hexagons. However,
the minimum energy Skyrmion with $B=13$ has $O$ symmetry and we have
also been able to find a Skyrmion with $O_h$ symmetry with $B=17$
which is a low-energy saddle point. Many such solutions in
which there is a four-valent bond connecting 4 pentagons can be
related to a fullerene polyhedron by symmetry enhancement as we shall
discuss in the next section. This produces an extra twist to the
fullerene hypothesis which allows octahedral Skyrmion solutions.

\subsection{Symmetry enhancement}
\label{sec-enhance}

\begin{figure}
\centerline{\epsfxsize=10cm\epsffile{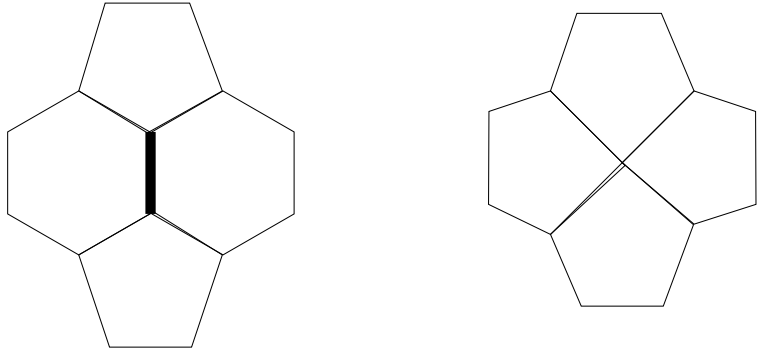}}
\caption{On the left is a configuration comprizing 
of two pentagons separated by two hexagons. Such configurations 
are prevalent in many fullerene polyhedra. On the right is what
 can be created by a single symmetry enhancement operation as
appears to take place for $B=9$ and $B=13$. It is usual for such an
operation to be accompanied by a similar operation on a diametrically
opposite face of the associated polyhedron.}    
\label{fig-enhance}
\end{figure}

We have seen that, in keeping with the fullerene hypothesis and the
expectations of ref.~\cite{BS2}, most links in the polyhedra associated
with Skyrmion solutions are trivalent. In particular, 
the baryon density isosurface of what we have identified as the
minimum energy solutions consist of a trivalent polyhedron for
all cases except $B=9$ and $B=13.$ In these two cases the polyhedra
contain four-valent vertices which 
means that they are not fullerene Skyrmions, since by definition
all links are trivalent. However, it turns out that these two
exceptional cases can be obtained from fullerenes by the application
of a simple rule, which we refer to as symmetry enhancement, and shall
now explain.

Consider part of a fullerene which has the form shown in the first
figure in  fig.~\ref{fig-enhance},
consisting of two pentagons and two hexagons with a
$C_2$ symmetry.  The symmetry enhancement operation is to shrink the
edge which is common to the two hexagons (the thick line) until it has
zero length, which results in the coalescence of two vertices.  The
final object formed is shown in the second figure in
fig.~\ref{fig-enhance}.  It has a
four-valent vertex connecting four pentagons and the symmetry is
enhanced from $C_2$ to $C_4.$ We find, empirically, that symmetry
enhancement operations appear to take place in pairs, with a particular
operation always being accompanied by the same operation on an
opposite face of the associated polyhedron.

There is a ${\rm C}_{28}$ fullerene with $D_2$ symmetry (denoted 28:1
in ref.\cite{atlas})
that contains two of the structures shown in the first figure in
fig.~\ref{fig-enhance}. 
If the above symmetry enhancement operation is performed on
both these structures then the resulting object is precisely the
$D_{4d}$ configuration of the $B=9$ Skyrmion described earlier. There
are also $D_2$ symmetric $C_{44}$ (denoted 44:75 and 44:89 in 
ref.~\cite{atlas}) fullerenes to which
similar statements apply.  In this case, which is $B=13$, there are an
equal number of  pentagons and hexagons (12 of each) and so a very
symmetric configuration can be obtained by applying the symmetry
enhancement operation at all possible vertices (in this case 6)
which results in the
cubic Skyrmion. Selective application of the symmetry enhancement rule
to these fullerenes allows one to create the associated polyhedron for
the $D_{4d}$ symmetric Skyrmion at this charge, and also that for a
$D_4$ symmetric Skyrmion, whose rational map we have been unable to
deduce since the minimum energy rational map in this class of maps
has $D_{4d}$ symmetry.

In the context of fullerenes it is, of course, impossible for vertices
to coalesce since they correspond to the positions of the carbon
atoms, but for Skyrmions the vertices represent local maxima of the
baryon density and so there is no restriction that they be distinct;
it just appears that in most cases it is energetically favourable to
have distinct vertices.
 Note that, by an examination of the baryon density isosurface by eye,
it can often be difficult to identify whether a given vertex is tri-valent
or four-valent, since the edge length required to be zero for
symmetry enhancement could be small, but non-zero. 

\subsection{Vertices, faces and rational maps}

We will now attempt to explain the various features of the Skyrmions we
have created by considering the basic properties of the rational map ansatz.
Recall that the baryon density of a Skyrmion is localized
around the edges of a polyhedron. The face centres of this
polyhedron are given by the zeros of the Wronskian of the numerator
and the denominator 
\be 
w(z)=p'(z)q(z)-q'(z)p(z)\,,
\label{wronskian}
\ee 
which is in general a degree $2B-2$ polynomial in $z.$
All the solutions which we have created have the
property that all the roots of (\ref{wronskian}) are distinct and
hence within the rational map ansatz this is a vindication of one of the
GEM rules, in that it explains why the number of faces of the 
polyhedron is $F=2B-2.$

Often (though not always, as we shall discuss further below) the
position of the 
vertices correspond to local maxima of the density which occur in the
integrand defining $\I$ in equation (\ref{i}).  This density depends
on the modulus of the rational map and its derivative so in general it
is not possible to obtain such a simple characterization of the
location of its maxima. However, in  particularly symmetric cases they can be
identified with the zeros of a polynomial constructed from the Hessian
of the Wronskian \cite{Con}. Explicitly, the polynomial is 
\be
H=(2B-2)w(z)w''(z)-(2B-3)w'(z)^2\,,
\label{hessian}
\ee 
and has degree $4(B-2).$ As an example in which the above formula does work,
consider the degree 7 rational map describing the icosahedrally
symmetric minimal energy charge 7 Skyrmion~\cite{HMS} 
\be
R=\frac{z^7-7z^5-7z^2-1}{z^7+7z^5-7z^2+1}\,.
\ee 
The Wronskian and the
Hessian in this case are given by \be w=28z(z^{10}+11z^5-1), \ \ \
H=-8624(z^{20}-228z^{15}+494z^{10}+228z^5+1)\,, \ee which are
proportional to the Klein polynomials associated with the faces and
vertices of a dodecahedron respectively \cite{Kl}.

Note that if the zeros of the Hessian (\ref{hessian}) could always be
identified with the vertices of the polyhedron then this would
explain another of the GEM rules, that is, the fact that the number of
vertices is equal to  $V=4(B-2).$ In this case the number of edges
$E=6(B-2)$ by the Euler formula, and hence $E=3V/2$, that is, the
polyhedron is trivalent.
Unfortunately, this is not true in
general, though for the minimal energy maps it may be the case that
the zeros of (\ref{hessian}) give good approximations to the positions
of the maxima.

We have already discussed explicit cases where the number of vertices
and edges is affected by symmetry enhancement, and therefore it comes
as no surprize that we cannot make general statements about the $E,F$
and $V$ based solely on the rational map ansatz.
Although we do not have a general global characterization of the vertices
it is possible, by a
 local analysis of the rational map, to check
whether a given point is a local maximum  and to obtain its
valency.\footnote{We thank Nick  Manton for suggesting this
possibility.}  By using the freedom to perform rotations of both the
domain and target spheres it is always possible to choose the point we
are considering to be given by $z=0$ and the rational map to have a
local expansion about this point of the form \be R=\alpha(z+\beta
z^{p+1}+O(z^{p+2}))
\label{local}
\ee where $\alpha$ and $\beta$ are real positive constants. A possible
exception to this is if the derivative of the map is zero at the point
we  are considering, but since such points correspond to face centres
they are clearly not maxima and may be ignored for our purposes here.
Substituting the
expansion (\ref{local}) into the expression for the baryon density one
can obtain the following result.

If $p>2$ then $z=0$ is a $p$-valent vertex if $\alpha>1.$ If $p=1$
then $z=0$ is not a vertex. The remaining case of $p=2$ is a little
more subtle.  In many cases there is a one-to-one correspondence
between the vertices and the  local maxima of the baryon density
polyhedron. However, this is not always true and in some cases (the
lowest charge example being $B=5$) only some of the local maxima are
vertices, whilst others correspond to midpoints of an edge.  In this
situation some of the edges may appear thicker than others, reflecting
their local maxima nature. The rational map description of such a
bivalent maximum is the final $p=2$ case, where a local maximum
requires the more  restrictive condition that $\alpha>\sqrt{1+3\beta}.$

As an example of this analysis, consider the $B=9$ map with  $D_{4d}$
symmetry  given
by (\ref{9a}). Expanding this map about the point $z=0$ gives \be
R=az+ib(1-a)z^5+...  \ee which can be rotated into the form
(\ref{local}) with $p=4$, $\alpha=-a =3.38>1$ and $\beta=-b(1-a).$
Thus the point $z=0$ is a four-valent vertex, as we have observed from
the baryon density plot.
The other minimizing map with four-valent vertices (this time six of
them) is the $B=13$ $O$ map (\ref{13_O}), which can be checked in a
similar way.

\subsection{Isomerism --- local minima and saddles}

We have argued strongly that there is an analogy between Skyrmion
solutions and polyhedra  found in carbon chemistry.
Moreover, at some charges we
have found more than one solution which has very low energy, and
therefore it might seem sensible within the analogy to chemistry 
to describe these solutions as
isomers, whether they be saddle point solutions or local
minima~\footnote{The existence of degenerate minima, would probably
require some kind of symmetry between the solutions. Although we
cannot rule it out, it appears to be very unlikely.}. 

In most cases, for example, $B=9,16,17,19$ and $22$ the symmetries and
structures of the known isomers are unrelated to those of the minimal
energy Skyrmions and in these cases 
it has been easy to identify the minimum energy configurations using
the relaxation of initially well-separated clusters. The cases of
$B=10$ and $13$ are interesting since there are known configurations
whose associated polyhedra are related by symmetry enhancement. We
have already commented that the polyhedron associated with $O$ symmetry
at $B=13$ can be created by 6 symmetry enhancement operations from a
$D_2$ fullerene polyhedron, and that with $D_{4d}$ symmetry 
requires just 2.

However, we have not discussed the $B=10$ solutions in this
context. The $D_{4d}$ and $D_{3h}$ solutions do not appear to be
related to this concept, but one can understand the $D_3$ and $D_{3d}$
solutions in terms of a more symmetric polyhedron which can be created
from either by 3 symmetry enhancement operations. Clearly, this highly
symmetric configuration, which is likely to be of higher energy,
can be thought of as a saddle point in configuration space, with the
two minima on either side. It is interesting to speculate that the
true minimum energy Skyrmion is more closely associated with a
polyhedron in which partial symmetry enhancement has taken place, that
is, the bond lengths have shrunk, but not totally to a four valent
bond. This might explain our difficulty in identifying the symmetry of
the true minima using our methods.

\subsection{Skyrmion architecture}

One of the most important reasons for performing full field
simulations in our study of Skyrmions is to verify that the minimal
energy Skyrmion (at least for $B\le 22$) consists of a single shell
structure, which is the main assumption in the rational map ansatz. In
this section we speculate on the kinds of structures  which may form
for Skyrmions of higher charge.

The lowest known value for the energy per baryon in the Skyrme model
arises from an infinite three-dimensional cubic crystal~\cite{CJJVJ},
with an energy ~\cite{BS3} only $3.6\%$ above the Faddeev-Bogomolny
bound.  In considering a large single-shell fullerene Skyrmion, where
hexagons are dominant, the twelve pentagons may be viewed as defects,
inserted into a flat hexagonal structure, in order to generate the
required curvature necessary to close the shell. Energetically the
optimum structure of this form is an infinite hexagonal lattice and
this was constructed in one of our earlier papers \cite{BS3}, and
found to have an energy per baryon which is $6.1\%$ above the
Faddeev-Bogomolny bound. This is therefore the value to which a bigger
and bigger single-shell structure will asymptote. Since this value is
higher than for the Skyrme crystal it is reasonable to expect that
above some critical charge $B^\star$,  the minimal energy Skyrmion
will resemble a portion cut from the crystal rather than a single
shell. However, it is very difficult to estimate the value of
$B^\star$ (note that we have seen that at least $B^\star>22$) since it
relies on a delicate comparison of the surface to volume energy of a
finite portion of the crystal and this is very sensitive to the way in
which the portion of the crystal is smoothed off at the edges.  As the
crystal is basically composed of stacking $B=4$ cubes together then
$B=32$ is the first charge at which any sizeable chunk of the crystal
can emerge, and even then it has a large area to volume ratio, so
perhaps the charge will need to be even larger than this before a
crystal regime takes over from the shell regime.

An intermediate between a single-shell and a crystal is a multi-shell
structure and this has recently been studied by Manton and Piette
\cite{MP}. For the charges they consider ($B=12,13,14$) the relaxation
of an initial multi-shell structure produces a single-shell
configuration which has relatively high energy in comparison with the
minimal energy single-shell Skyrmions we have found.  As a shell may
be thought of as a spherical domain wall, connecting the two vacua
$U=\pm 1$, then all configurations with an odd number of shells have
$U=-1$ at the centre, whereas if there are an even number of shells
then $U=1$ at the centre. Thus a single shell structure obtained from
an initial odd number of shells can relax to one of the configurations
we have found. In fact, as we have already mentioned, the $B=13$
Skyrmion obtained by Manton and Piette is the $O_h$ symmetric saddle
point solution whereas the minimal energy Skyrmion is only $O$
symmetric. 

In summary, there are a number of alternatives to a single-shell
structure for higher charge Skyrmions and what is remarkable is that
none of these alternatives appear to arise at least for $B\le 22.$ It
seems reasonably clear that single-shells can not be the whole story
for large enough charge, but whether this charge is so large as to be
irrelevant in applications to nuclear physics has yet to be determined.

\subsection{Relation to applications}
\label{sec-app}

In the introduction we commented on two diverse motivations for
creating Skyrmion solutions, namely 
from a purely mathematical point of view, to study an interesting
class of maps between 3-spheres which generalize the harmonic map
equations, and from a physical perspective to investigate a
phenomenological model of nuclei. Here, we comment briefly on the
relevance of our results to these two applications and suggest
interesting avenues for future research which we have opened up with
this work.

We have already noted that the Skyrme model is the simplest model in
which one finds stable solitonic solutions which correspond to maps
from $S^3$ to $S^3$, and so our solutions may have some generality to
other extensions. An interesting feature of the solutions which
we have found is that, in some sense, they can be thought of as being close to a
conformal map between the two 3-spheres, for which the
three eigenvalues of the strain tensor, $\lambda_1^2,\lambda_2^2,\lambda_3^2$,
would all be equal. The rational map ansatz, which we have seen provides a good
approximation to the true solutions, has two of these eigenvalues equal
and it has been observed~\cite{krusch} that the shape of the profile function appears to
be such that the deviation from a conformal map is minimized
when averaged over space.

For a conformal map which is locally an isometry the
  values of ${\cal E}$ and ${\cal B}$ would be exactly equal and so any deviations from the map
being locally isometric can be visualized by plotting ${\cal E}-{\cal
B}$. When one does this the relevant isosurface is highly localized around the
edges of the associated polyhedron and also in the centre of each face
where ${\cal B}$ is close to zero and ${\cal E}$ is large in comparison.
Such an isosurface is a plot of second order effects due to curvature.
The fact that the associated polyhedra are generally of the
fullerene type is also interesting because in chemistry such structures
arise since they minimize what is called steric strain, 
the overall strain of the delocalized electron distribution. Using
this analogy we suggest that the effects of the strain tensor for maps
between 3-spheres can be thought of as being analogous to steric
strain in fullerene molecules.

Finally, we should note that the existence of a Skyrmion with a
particular symmetry, which can be described by a rational map,
 implies that there also exists an SU(2) BPS monopole
with the same symmetry, although, of course, all BPS monopoles of a 
given charge have the same energy.
The fields and Lagrangians of monopoles and Skyrmions are very different
but the structures which arise in each case are remarkably similar.
This suggests that these types of configurations may be generic as low
energy states in a variety of 3-dimensional soliton models and elsewhere.

\begin{figure}
\centerline{\epsfxsize=10cm\epsfysize=10cm\epsffile{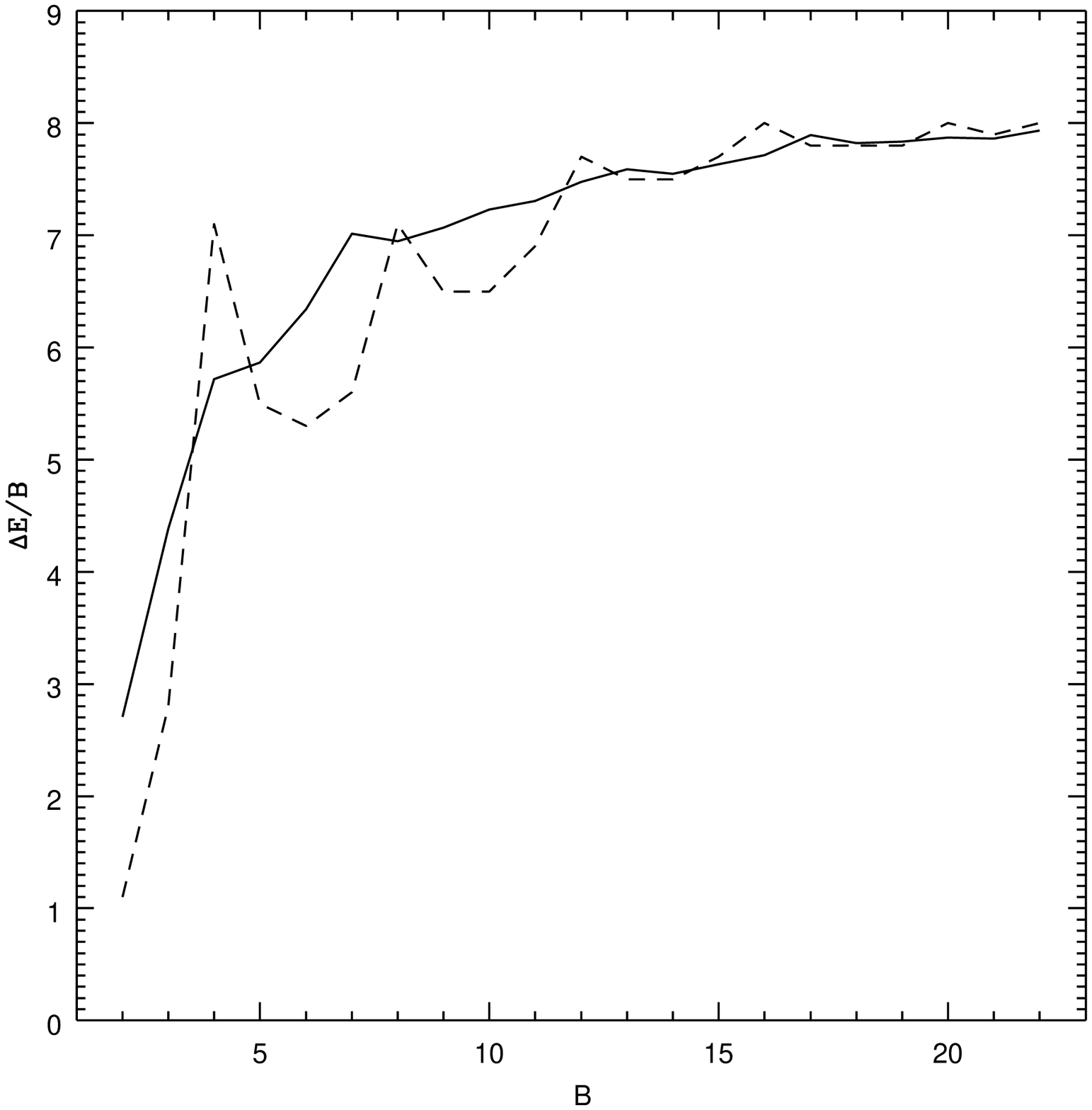}}
\caption{The binding energy per baryon number for Skyrmions (solid line) compared,
using an arbitrary scaling, to the binding energy per nucleon of real
nuclei (dashed line). The main thing that one should note is that the approximate
shape, an increase to a plateau, is present in both.}    
\label{fig-comp}
\end{figure}

Although the original motivation for studying the Skyrme model was 
to make quantitative predictions for the properties of nuclei based
on a model which is derived in some limit from QCD, this has
historically been a tricky business. Our hope, based on the extensive
results that we have presented here, is that some progress can be
made in this direction. Part of the problem in achieving such a goal
is how one should understand the model in the context of nuclei. Based
on the idea that it is a low energy effective action for QCD several
studies have
attempted to quantize Skyrmion solutions as rigidly rotating spinning
tops for $B=1,2$ and $3$ (see, for
example, refs.\cite{ANW,Car1,Car2}) quantifying
corrections in terms of their position in the $1/N_{\rm c}$ expansion.
This is not only complicated but probably is also too simplified an approach,
as demonstrated by the more sophisticated quantization
of the $B=2$ Skyrmion in ref.\cite{LMS}.
Here we would like to make a few interesting
points in terms of thinking of the solitons as a phenomenological
model for nuclei which incorporates classical isospin.

The first thing that we would like to discuss is the shell structure
of the soliton solutions which we have created. Naively, one might
think that such a structure is incompatible with the solutions
modelling nuclei which are usually assumed to be solid lumps.
What one has to appreciate to reconcile this is, that until the
solutions are quantized, the charge can be thought of as a fluid. If a
nuclei were assumed to be comprized of a simple positively charge
fluid then a shell structure would be expected under the action of the
strong force since there is a long range, but weak, attraction at long
distances, and a short distance repulsion due to nucleon-nucleon
repulsion. Therefore, the hollow shell structure of the
multi-Skyrmions which we have observed is largely due to the continuum
version of nucleon-nucleon repulsion.

Following on from this point we should note that some of the
features of the classical values of $I_B$, the ionization energy, are
very much in line with expectations based on nuclei.
 Let us focus in detail on the solutions for $B=4$ and $B=5$. The $B=4$
solution is highly symmetric and $I_4$ is relatively large, whereas
the $B=5$ solution has little symmetry and $I_5$ is much
smaller. This is exactly as one might have expected since $^4{\rm He}$
is the most stable nucleus whereas there is no naturally occurring
stable nucleus with $A=5$. Since the packing structure of the
solutions is a feature of the symmetry of the solution
this suggests that there may be something even more than just a
good model of the strong force potential within the Skyrme model.

There is also an interesting trend in $\Delta E/B$, the classical
binding energy per baryon, which appears to asymptote to a value
defined by that of an infinite Skyrmion lattice. We have plotted
this compared to the experimentally determined values for nuclei with
$A=1$ to $22$~\cite{CG,HGG} in fig.~\ref{fig-comp} with an arbitrary
normalization factor (which amounts to multiplying the curve in
fig.~\ref{fig-bind} by about 50) accounting for our ability to define
the Skyrmion energy units. Although this is crude, it makes the
point that the curve has the correct shape. This is very
encouraging and is the subject of on-going research.

Since the fullerene polyhedra are clearly very important for our
understanding of Skyrmions, and as we have just argued there
are a number of appealing features of the model for explaining the
properties of nuclei, it is tempting to make an analogy between the
delocalized electron distributions in fullerenes and nuclear charge
distributions. Although the analogy is not exact, it might be possible
to relate the Skyrme model to density functional theory methods (see,
for example, ref.~\cite{DFT}) used
in the study of electron distributions. 

\section{Conclusion}\news

We have performed an exhaustive study of minimal energy Skyrmions
for all charges upto $B=22$, using a variety of methods and involving
a substantial amount of CPU time on a parallel machine.
At each charge we have discussed in detail the symmetry, structure
and energy of the minimal energy Skyrmion (and often several others)
in addition to providing an approximate description by presenting
its associated rational map. Supplementary to the detailed investigation
at specific charges we have found a number of interesting general phenomena.
These include the verification of the fullerene hypothesis, which applies
to all except two cases (which can be understood in terms of symmetry
enhancement), the discovery that there are often several Skyrmions with
very different symmetries from the minimal one but nonetheless have energies
which are remarkably close to the minimal value, and finally the confirmation
that the shell-like structure of Skyrmions continues to large charges (at 
least $B=22$) with the rational map ansatz providing an effective approximation
to the true solution. Hopefully this comprehensive piece of work will 
provide a useful foundation for further studies on Skyrmions, 
both mathematical and physical, with the ultimate
aim being a comparison with experimental data on nuclei.

\section*{Acknowledgements}
\news
Many thanks to Conor Houghton, Nick Manton and Tom Weidig for useful
discussions.  We thank the EPSRC (PMS) and PPARC (RAB) for Advanced
Research Fellowships. 
PMS acknowledges the EPSRC for the grant GR/M57521.
The parallel computations were performed on the
COSMOS at the National Cosmology Supercomputing Centre in Cambridge.

\def\jnl#1#2#3#4#5#6{\hang{#1, {\it #4\/} {\bf #5}, #6 (#2).} }
\def\jnltwo#1#2#3#4#5#6#7#8{\hang{#1, {\it #4\/} {\bf #5}, #6; {\it
ibid} {\bf #7} #8 (#2).} } \def\prep#1#2#3#4{\hang{#1, #4.} }
\def\proc#1#2#3#4#5#6{{#1 [#2], in {\it #4\/}, #5, eds.\ (#6).} }
\def\prep#1#2#3#4{\hang{#1, #4 (#2).}}
\def\book#1#2#3#4{\hang{#1, {\it #3\/} (#4, #2).} }
\def\jnlerr#1#2#3#4#5#6#7#8{\hang{#1 [#2], {\it #4\/} {\bf #5}, #6.
{Erratum:} {\it #4\/} {\bf #7}, #8.} } \def\prl{Phys.\ Rev.\ Lett.}
\def\pr{Phys.\ Rev.}  \def\pl{Phys.\ Lett.}  \def\np{Nucl.\ Phys.}
\def\prp{Phys.\ Rep.}  \def\rmp{Rev.\ Mod.\ Phys.}  \def\cmp{Comm.\
Math.\ Phys.}  \def\mpl{Mod.\ Phys.\ Lett.}  \def\apj{Astrophys.\ J.}
\def\apjl{Ap.\ J.\ Lett.}  \def\aap{Astron.\ Ap.}  \def\cqg{Class.\
Quant.\ Grav.}  \def\grg{Gen.\ Rel.\ Grav.}  \def\mn{Mon.\ Not.\ Roy.\
Astro.\ Soc.}
\def\ptp{Prog.\ Theor.\ Phys.}  \def\jetp{Sov.\ Phys.\ JETP}
\def\jetpl{JETP Lett.}  \def\jmp{J.\ Math.\ Phys.}  \def\zpc{Z.\
Phys.\ C} \def\cupress{Cambridge University Press} \def\pup{Princeton
University Press} \def\wss{World Scientific, Singapore}
\def\oup{Oxford University Press}

\end{document}